# Orbits and Manifolds near the Equilibrium Points around a Rotating Asteroid

Yu Jiang[1, 2], Hexi Baoyin[2], Junfeng Li[2]

1. State Key Laboratory of Astronautic Dynamics, Xi'an Satellite Control Center, Xi'an 710043, China

2. School of Aerospace, Tsinghua University, Beijing 100084, China

Y. Jiang (✉) e-mail: jiangyu_xian_china@163.com (corresponding author)

H. Baoyin (✉) e-mail: baoyin@tsinghua.edu.cn (corresponding author)

**Abstract.** We study the orbits and manifolds near the equilibrium points of a rotating asteroid. The linearised equations of motion relative to the equilibrium points in the gravitational field of a rotating asteroid, the characteristic equation and the stable conditions of the equilibrium points are derived and discussed. First, a new metric is presented to link the orbit and the geodesic of the smooth manifold. Then, using the eigenvalues of the characteristic equation, the equilibrium points are classified into 8 cases. A theorem is presented and proved to describe the structure of the submanifold as well as the stable and unstable behaviours of a massless particle near the equilibrium points. The linearly stable, the non-resonant unstable, and the resonant equilibrium points are discussed. There are three families of periodic orbits and four families of quasi-periodic orbits near the linearly stable equilibrium point. For the non-resonant unstable equilibrium points, there are four cases; for the periodic orbit and the quasi-periodic orbit, the structures of the submanifold and the subspace near the equilibrium points are studied for each case. For the resonant equilibrium points, the dimension of the resonant manifold is greater than four, and we find at least 1 family of periodic orbits near the resonant equilibrium points. Besides, this theory is applied to asteroids 216 Kleopatra, 1620 Geographos, 4769 Castalia, and 6489 Golevka.

# 1. Introduction

Recently, several sample return missions to Near-Earth Asteroid (NEA) are selected (Barucci 2012; Barucci et al. 2012a; Barucci et al. 2012b; Boynton et al. 2012; Brucato 2012; Duffard et al. 2011), including MarcoPolo-R, OSIRIS-Rex, etc. MarcoPolo-R is a sample return mission to a primitive Near-Earth Asteroid (NEA) selected for the assessment study in the framework of ESA Cosmic Vision 2015-25 program (Barucci 2012; Barucci et al. 2012a; Barucci et al. 2012b; Brucato 2012). The baseline target of MarcoPolo-R is a binary asteroid (175706) 1996 FG3, which offers a very efficient operational and technical mission profile (Brucato 2012). OSIRIS-Rex is an asteroid sample-return mission that was selected by NASA in May 2011 as the third New Frontiers Mission (Boynton et al. 2012).

MarcoPolo-R can help one to understand the unique geomorphology, dynamics and evolution of a binary NEA (Barucci et al. 2012a). These missions make the dynamics of a spacecraft (which is modeled as a massless particle hereafter) near asteroids becomes an interesting topic. By applying the classic method of Legendre polynomial series to the gravity potential of asteroids, the gravity potential may diverge at some points in the gravitational field due to the irregular shape of the asteroid (Balmino 1994; Elipe & Lara 2003). Several methods have been considered to solve this difficulty. Some simple-shaped bodies have been used to simulate the motion of a particle moving near an asteroid as well as to help understanding the equilibrium points and the periodic orbits near asteroids. The dynamics around a straight segment (Riaguas et al. 1999; Elipe & Lara 2003), a solid circular ring (Broucke & Elipe 2005), a homogeneous annulus disk (Angelo & Claudio 2007; Fukushima 2010), and a homogeneous cube (Liu et al. 2011) have been studied in detail. These studies are limited to simple and symmetric bodies, which rotate around their symmetric axes. For these bodies, the characteristic equation at the equilibrium points is a quartic equation (Liu et al. 2011). Werner & Scheeres (1996) modelled the shape and the gravity potential of the asteroids using homogeneous polyhedrons and applied this method to 4769 Castalia. The polyhedron model is more precise than the simple-shaped-body models and the Legendre polynomial model. However, this model presents many free parameters (Elipe & Lara 2003; Werner & Scheeres 1996).

The dynamic equation of a massless particle near a rotating asteroid in the body-fixed frame was studied and applied to 4769 Castalia and 4179 Toutatis by Scheeres et al. (1996, 1998). The zero-velocity surface is defined using the Jacobi integral, and this surface separates the forbidden regions from the allowable regions for the particle (Scheeres et al. 1996; Yu & Baoyin 2012a). The periodic orbit families near the asteroids can be calculated using Poincaré maps and numerical iterations (Scheeres et al. 1996, 1998; Yu & Baoyin 2012b). The positions of the equilibrium points near the asteroids as well as the eigenvalues and the periodic orbits near the equilibrium points can be numerically calculated (Yu & Baoyin 2012a).

The stability of the equilibrium points was discussed by several authors using special models. The periodic orbits of a particle around a massive straight segment were investigated in Riaguas et al. (1999). A finite straight segment with constant linear mass density was considered to study the chaotic motion and the equilibrium points around the asteroid (433) Eros (Elipe & Lara 2003). The dynamics of a particle around a homogeneous annulus disk was analysed in Alberti & Vidal (2007). Moreover, a precise numerical method to calculate the acceleration vector that is caused by a uniform ring or disk was described in Fukushima et al. (2010). The dynamics of a particle around a non-rotating 2nd-degree and order-gravity field was examined, and the closed-form solutions for the averaged orbital motion of the particle were derived (Scheeres & Hu 2001).

This paper aims to derive the characteristic equation of the equilibrium points near a rotating asteroid and to provide some corollaries about the stability, the instability, and the resonance of the equilibrium points. Then, we will discuss the orbits and the structure of the submanifold and the subspace near the equilibrium points. This work can reveal the dynamics near the equilibrium points of a rotating asteroid while providing the theoretical basis for orbital design and control of spacecraft motion near an asteroid.

The linearised equations of motion that are relative to the equilibrium points in the gravitational field of a rotating asteroid are derived in section 2; then, the characteristic equation of the equilibrium points is presented. The stability conditions of the equilibrium points in the potential field of a rotating asteroid are obtained and proved in section 3. Furthermore, a new metric in a smooth manifold is provided in Section 4 to link the orbit and the geodesic of the smooth manifold. The equilibrium points are classified into eight cases using the eigenvalues of the characteristic equation in section 5. It is found that the structure of the submanifold near the equilibrium point is related to the eigenvalues. A theorem that describes the structure of the submanifold and the stable and unstable behaviours of the particle near the equilibrium points is proved. Using this theorem, the linearly stable, the non-resonant unstable and the resonant equilibrium points are studied. The equilibrium point is resonant if and only if at least two pure imaginary eigenvalues are equal. Subsequently, four families of quasi-periodic orbits are found near the linearly stable equilibrium point. The periodic orbit, the quasi-periodic orbit, the structure of the submanifold and the subspace near the non-resonant unstable equilibrium points are studied. For the resonant equilibrium points, there are three cases. For each case, the resonant orbits and the resonant manifolds are analysed. There is at least one family of periodic orbits near each resonant equilibrium point, and the dimension of the resonant manifold is greater than four.

The theory is applied to asteroids 216 Kleopatra, 1620 Geographos, 4769 Castalia, and 6489 Golevka. For the asteroid 216 Kleopatra, 1620 Geographos, and 4769 Castalia, the equilibrium points are unstable, which are denoted as E1, E2, E3, and E4. There are two families of periodic orbits and one family of quasi-periodic orbits on the central manifold near each of the equilibrium points E1 & E2, while there is only one family of periodic orbits on the central manifold near each of the other two equilibrium points E3 & E4.

For the asteroid 6489 Golevka, two of the equilibrium points are unstable, which are denoted as E1 and E2; while the other two of the equilibrium points are linearly stable, which are denoted as E3 and E4. There are two families of periodic orbits as well as one family of quasi-periodic orbits on the central manifold near each of the two unstable equilibrium points E1 and E2, while there are three families of periodic orbits as well as four families of quasi-periodic orbits on the central manifold near each of the two linearly stable equilibrium points E3 and E4.

## 2. Equations of Motion

The equation of motion relative to the rotating asteroid can be given as (Scheeres et al. 1996)

$$\ddot{\mathbf{r}} + 2\boldsymbol{\omega} \times \dot{\mathbf{r}} + \boldsymbol{\omega} \times (\boldsymbol{\omega} \times \mathbf{r}) + \dot{\boldsymbol{\omega}} \times \mathbf{r} + \frac{\partial U(\mathbf{r})}{\partial \mathbf{r}} = 0, \qquad (1)$$

where $\mathbf{r}$ is the body-fixed vector from the centre of mass of the asteroid to the particle, $\boldsymbol{\omega}$ is the rotational angular velocity vector of the asteroid relative to the inertial frame of reference, and $U(\mathbf{r})$ is the gravitational potential.

Let us define a function $H$ as (Scheeres et al. 1996)

$$H = \frac{1}{2}\dot{\mathbf{r}} \cdot \dot{\mathbf{r}} - \frac{1}{2}(\boldsymbol{\omega} \times \mathbf{r})(\boldsymbol{\omega} \times \mathbf{r}) + U(\mathbf{r}) \qquad (2)$$

If $\boldsymbol{\omega}$ is time-invariant, then $H$ is also time-invariant, and it is called the Jacobian constant.

The effective potential can be defined as (Scheeres et al. 1996; Yu & Baoyin

2012a)

$$V(\mathbf{r}) = -\frac{1}{2}(\boldsymbol{\omega} \times \mathbf{r})(\boldsymbol{\omega} \times \mathbf{r}) + U(\mathbf{r}) \tag{3}$$

Hence, the equation of motion can be written as

$$\ddot{\mathbf{r}} + 2\boldsymbol{\omega} \times \dot{\mathbf{r}} + \dot{\boldsymbol{\omega}} \times \mathbf{r} + \frac{\partial V(\mathbf{r})}{\partial \mathbf{r}} = 0 \tag{4}$$

For a uniformly rotating asteroid, equation (4) can be written as (Yu & Baoyin 2012a)

$$\ddot{\mathbf{r}} + 2\boldsymbol{\omega} \times \dot{\mathbf{r}} + \frac{\partial V(\mathbf{r})}{\partial \mathbf{r}} = 0, \tag{5}$$

and the Jacobian constant can be given by

$$H = \frac{1}{2}\dot{\mathbf{r}} \cdot \dot{\mathbf{r}} + V(\mathbf{r}) \tag{6}$$

The zero-velocity manifolds are determined by the following equation (Scheeres et al. 1996; Yu & Baoyin 2012a)

$$V(\mathbf{r}) = H \tag{7}$$

The inequality $V(\mathbf{r}) > H$ denotes the forbidden region for the particle, whereas the inequality $V(\mathbf{r}) \leq H$ denotes the allowable region for the particle. The equation $V(\mathbf{r}) = H$ implies that the velocity of the particle relative to the rotating body-fixed frame is zero.

Let $\omega$ be the modulus of the vector $\boldsymbol{\omega}$; then, the unit vector $\mathbf{e}_z$ is defined by $\boldsymbol{\omega} = \omega \mathbf{e}_z$. The body-fixed frame is defined through a set of orthonormal right-hand unit vectors $\mathbf{e}$

$$\mathbf{e} \equiv \begin{Bmatrix} \mathbf{e}_x \\ \mathbf{e}_y \\ \mathbf{e}_z \end{Bmatrix} \tag{8}$$

The equilibrium points are the critical points of the effective potential $V(\mathbf{r})$. Thus, the equilibrium points satisfy the following condition

$$\frac{\partial V(x,y,z)}{\partial x} = \frac{\partial V(x,y,z)}{\partial y} = \frac{\partial V(x,y,z)}{\partial z} = 0, \qquad (9)$$

where $(x, y, z)$ are the components of $\mathbf{r}$ in the body-fixed coordinate system. Let $(x_L, y_L, z_L)^T$ denote the coordinates of a critical point; the effective potential $V(x, y, z)$ can be expanded using Taylor expansion at the equilibrium point $(x_L, y_L, z_L)^T$. To study the stability of the equilibrium point, the equations of motion relative to the equilibrium point are linearised, and the characteristic equation of motion is derived. In addition, it is necessary to check whether any solutions of the characteristic equation have positive real components. The Taylor expansion of the effective potential $V(x, y, z)$ at the point $(x_L, y_L, z_L)^T$ can be written as

$$V(x,y,z) = V(x_L, y_L, z_L) + \frac{1}{2}\left(\frac{\partial^2 V}{\partial x^2}\right)_L (x-x_L)^2 + \frac{1}{2}\left(\frac{\partial^2 V}{\partial y^2}\right)_L (y-y_L)^2 + \frac{1}{2}\left(\frac{\partial^2 V}{\partial z^2}\right)_L (z-z_L)^2$$
$$+ \left(\frac{\partial^2 V}{\partial x \partial y}\right)_L (x-x_L)(y-y_L) + \left(\frac{\partial^2 V}{\partial x \partial z}\right)_L (x-x_L)(z-z_L) + \left(\frac{\partial^2 V}{\partial y \partial z}\right)_L (y-y_L)(z-z_L) + \cdots$$

(10)

where the solving derivation sequence is changeable, which implies that $\frac{\partial^2 V}{\partial x \partial y} = \frac{\partial^2 V}{\partial y \partial x}$, $\frac{\partial^2 V}{\partial x \partial z} = \frac{\partial^2 V}{\partial z \partial x}$, $\frac{\partial^2 V}{\partial y \partial z} = \frac{\partial^2 V}{\partial z \partial y}$.

Let's define

$$\begin{array}{ll} & V_{xx} = \left(\frac{\partial^2 V}{\partial x^2}\right)_L \qquad V_{xy} = \left(\frac{\partial^2 V}{\partial x \partial y}\right)_L \\ \xi = x - x_L & \\ \eta = y - y_L, \quad V_{yy} = \left(\frac{\partial^2 V}{\partial y^2}\right)_L \quad \text{and} \quad V_{yz} = \left(\frac{\partial^2 V}{\partial y \partial z}\right)_L \qquad (11) \\ \zeta = z - z_L & \\ & V_{zz} = \left(\frac{\partial^2 V}{\partial z^2}\right)_L \qquad V_{xz} = \left(\frac{\partial^2 V}{\partial x \partial z}\right)_L \end{array}$$

Combining these equations with Eq. (5), the linearised equations of motion relative to the equilibrium point can be expressed as

$$\ddot{\xi} - 2\omega\dot{\eta} + V_{xx}\xi + V_{xy}\eta + V_{xz}\zeta = 0$$
$$\ddot{\eta} + 2\omega\dot{\xi} + V_{xy}\xi + V_{yy}\eta + V_{yz}\zeta = 0 \tag{12}$$
$$\ddot{\zeta} + V_{xz}\xi + V_{yz}\eta + V_{zz}\zeta = 0$$

It is natural to write the characteristic equation in the following form

$$\begin{vmatrix} \lambda^2 + V_{xx} & -2\omega\lambda + V_{xy} & V_{xz} \\ 2\omega\lambda + V_{xy} & \lambda^2 + V_{yy} & V_{yz} \\ V_{xz} & V_{yz} & \lambda^2 + V_{zz} \end{vmatrix} = 0 \tag{13}$$

Furthermore, it can be written as

$$\lambda^6 + \left(V_{xx} + V_{yy} + V_{zz} + 4\omega^2\right)\lambda^4 + \left(V_{xx}V_{yy} + V_{yy}V_{zz} + V_{zz}V_{xx} - V_{xy}^2 - V_{yz}^2 - V_{xz}^2 + 4\omega^2 V_{zz}\right)\lambda^2$$
$$+ \left(V_{xx}V_{yy}V_{zz} + 2V_{xy}V_{yz}V_{xz} - V_{xx}V_{yz}^2 - V_{yy}V_{xz}^2 - V_{zz}V_{xy}^2\right) = 0$$

(14)

where $\lambda$ denotes the eigenvalues of Eq. (12). Eq. (14) is a sextic equation for $\lambda$. The stability of the equilibrium point is determined by six roots of Eq. (13). Let $\lambda_i (i = 1, 2, \cdots, 6)$ be the roots of Eq. (13). The equilibrium point is asymptotically stable if and only if $\text{Re}\,\lambda_i < 0$ for $i = 1, 2, \cdots, 6$, i.e., the equilibrium point is a sink (minimum) of the nonlinear dynamics system (5). Furthermore, the equilibrium point is unstable if and only if there is a $i$ equals to one of $1, 2, \cdots, 6$, such that $\text{Re}\,\lambda_i < 0$, i.e., the equilibrium point is a source or a saddle.

## 3. Stability of the Equilibrium Points in the Potential Field of a Rotating Asteroid

In this section, the stability of the equilibrium points in the potential field of a rotating asteroid is studied. First, the roots of the characteristic equation at the equilibrium point are investigated; then, a theorem about the eigenvalues is given. In addition, a sufficient condition for the stability of equilibrium points is presented, which only depends on the Hessian matrix of the effective potential. Finally, a necessary and sufficient condition for the stability of equilibrium points is also presented.

Let's denote

$$P(\lambda) = \lambda^6 + \left(V_{xx} + V_{yy} + V_{zz} + 4\omega^2\right)\lambda^4 + \left(V_{xx}V_{yy} + V_{yy}V_{zz} + V_{zz}V_{xx} - V_{xy}^2 - V_{yz}^2 - V_{xz}^2 + 4\omega^2 V_{zz}\right)\lambda^2$$
$$+ \left(V_{xx}V_{yy}V_{zz} + 2V_{xy}V_{yz}V_{xz} - V_{xx}V_{yz}^2 - V_{yy}V_{xz}^2 - V_{zz}V_{xy}^2\right)$$

, and it follows that $P(\lambda) = P(-\lambda)$. Hence:

**Proposition 1**. If $\lambda$ is an eigenvalue of the equilibrium point in the potential field of a uniformly rotating asteroid, then $-\lambda$, $\bar{\lambda}$, and $-\bar{\lambda}$ are also eigenvalues of the equilibrium point. Namely, all eigenvalues are likely to have the form $\pm\alpha (\alpha \in \mathrm{R}, \alpha > 0)$, $\pm i\beta (\beta \in \mathrm{R}, \beta > 0)$, and $\pm\sigma \pm i\tau (\sigma, \tau \in \mathrm{R}; \sigma, \tau > 0)$. □

**Theorem 1**. If the matrix $\begin{pmatrix} V_{xx} & V_{xy} & V_{xz} \\ V_{xy} & V_{yy} & V_{yz} \\ V_{xz} & V_{yz} & V_{zz} \end{pmatrix}$ is positive definite, the equilibrium point in the potential field of a rotating asteroid is stable.

**Proof:**

Eq. (12) can be expressed as

$$\mathbf{M\ddot{X}} + \mathbf{G\dot{X}} + \mathbf{KX} = 0, \qquad (15)$$

where

$$\mathbf{X} = \begin{bmatrix} \xi & \eta & \zeta \end{bmatrix}^{\mathrm{T}}, \quad \mathbf{M} = \begin{pmatrix} 1 & 0 & 0 \\ 0 & 1 & 0 \\ 0 & 0 & 1 \end{pmatrix}, \quad \mathbf{G} = \begin{pmatrix} 0 & -2\omega & 0 \\ 2\omega & 0 & 0 \\ 0 & 0 & 0 \end{pmatrix}, \quad \mathbf{K} = \begin{pmatrix} V_{xx} & V_{xy} & V_{xz} \\ V_{xy} & V_{yy} & V_{yz} \\ V_{xz} & V_{yz} & V_{zz} \end{pmatrix}$$

The matrices $\mathbf{M}$, $\mathbf{G}$, and $\mathbf{K}$ satisfy $\mathbf{M}^{\mathrm{T}} = \mathbf{M} > 0$, $\mathbf{G}^{\mathrm{T}} = -\mathbf{G}$, and $\mathbf{K}^{\mathrm{T}} = \mathbf{K}$, respectively. The system that can be expressed by the equation $\mathbf{M}\ddot{\mathbf{X}} + \mathbf{G}\dot{\mathbf{X}} + \mathbf{K}\mathbf{X} = 0$ is stable if the matrix $\mathbf{K} = \begin{pmatrix} V_{xx} & V_{xy} & V_{xz} \\ V_{xy} & V_{yy} & V_{yz} \\ V_{xz} & V_{yz} & V_{zz} \end{pmatrix}$ is positive definite. □

Theorem 1 is a sufficient condition for the stability of the equilibrium points in the potential field of a rotating asteroid. The following theorem 2 will provide a necessary and sufficient condition for the stability of the equilibrium points in the potential field of a rotating asteroid.

**Theorem 2**. The equilibrium point in the potential field of a rotating asteroid is stable if and only if

$$\begin{cases} V_{xx} + V_{yy} + V_{zz} + 4\omega^2 > 0 \\ V_{xx}V_{yy} + V_{yy}V_{zz} + V_{zz}V_{xx} + 4\omega^2 V_{zz} > V_{xy}^2 + V_{yz}^2 + V_{xz}^2 \\ V_{xx}V_{yy}V_{zz} + 2V_{xy}V_{yz}V_{xz} > V_{xx}V_{yz}^2 + V_{yy}V_{xz}^2 + V_{zz}V_{xy}^2 \\ A^2 + 18ABC > 4A^3C + 4B^3 + 27C^2 \end{cases}, \tag{16}$$

where

$$\begin{cases} A = V_{xx} + V_{yy} + V_{zz} + 4\omega^2 \\ B = V_{xx}V_{yy} + V_{yy}V_{zz} + V_{zz}V_{xx} - V_{xy}^2 - V_{yz}^2 - V_{xz}^2 + 4\omega^2 V_{zz} \\ C = V_{xx}V_{yy}V_{zz} + 2V_{xy}V_{yz}V_{xz} - V_{xx}V_{yz}^2 - V_{yy}V_{xz}^2 - V_{zz}V_{xy}^2 \end{cases}$$

**Proof:**

Consider the linearised equation of motion relative to the equilibrium point

$$\mathbf{M}\ddot{\mathbf{X}} + \mathbf{G}\dot{\mathbf{X}} + \mathbf{K}\mathbf{X} = 0$$

Because the matrices $\mathbf{M}$, $\mathbf{G}$, $\mathbf{K}$ satisfy $\mathbf{M}^{\mathrm{T}} = \mathbf{M} > 0$, $\mathbf{G}^{\mathrm{T}} = -\mathbf{G}$, $\mathbf{K}^{\mathrm{T}} = \mathbf{K}$, The linearised equation is a gyroscope conservative system.

The characteristic equation of the gyroscope conservative system is

$$P(\lambda) = \lambda^6 + \left(V_{xx} + V_{yy} + V_{zz} + 4\omega^2\right)\lambda^4 + \left(V_{xx}V_{yy} + V_{yy}V_{zz} + V_{zz}V_{xx} - V_{xy}^2 - V_{yz}^2 - V_{xz}^2 + 4\omega^2 V_{zz}\right)\lambda^2$$
$$+ \left(V_{xx}V_{yy}V_{zz} + 2V_{xy}V_{yz}V_{xz} - V_{xx}V_{yz}^2 - V_{yy}V_{xz}^2 - V_{zz}V_{xy}^2\right) = 0$$

Using the stability condition of the gyroscope conservative system (Hughes 1986), the conclusion can easily be obtained. □

From Theorem 2, it can be noted that the necessary and sufficient condition for the stability of the equilibrium points in the potential field of a rotating asteroid depends on the effective potential and the rotational angular velocity of the asteroid.

## 4. Orbits, Differentiable Manifold and Geodesic

Previously, we denoted $(x, y, z)$ as the position of the spacecraft in the asteroid body-fixed coordinate system. Assuming that $\mathbf{A}^3$ is the topological space generated by $(x, y, z)$, the open sets of $\mathbf{A}^3$ are naturally defined. It is clear that $\mathbf{A}^3$ is a smooth manifold. In this section, a metric $d\rho^2$ is provided so that the geodesic of $\left(\mathbf{A}^3, d\rho^2\right)$ is the orbit of the particle relative to the asteroid.

**Theorem 3.** Denote $d\rho^2 = 2m(h-U)\left[d\mathbf{r}\cdot d\mathbf{r} - (\boldsymbol{\omega}\times\mathbf{r})\cdot(\boldsymbol{\omega}\times\mathbf{r})\right]$; then, the orbit of the particle relative to the asteroid on the manifold $H = h$ is the geodesic of $\left(\mathbf{A}^3, d\rho^2\right)$ with the metric $d\rho^2$. In addition, the geodesic of $\left(\mathbf{A}^3, d\rho^2\right)$ with the metric $d\rho^2$ is the orbit of the particle relative to the asteroid on the manifold $H = h$.

**Proof:**

The equation of motion is Eq. (1), and the Jacobian constant is expressed by Eq.

(2).

Let

$$J = \int_{t_1}^{t_2} \left[ d\mathbf{r} \cdot d\mathbf{r} - (\boldsymbol{\omega} \times \mathbf{r}) \cdot (\boldsymbol{\omega} \times \mathbf{r}) \right] dt \tag{17}$$

Then, on the manifold $H = h$, the variation is zero.

$$\delta J = 0$$

On the manifold $H = h$, the following equation holds

$$\frac{d\mathbf{r}}{dt} \cdot \frac{d\mathbf{r}}{dt} - (\boldsymbol{\omega} \times \mathbf{r}) \cdot (\boldsymbol{\omega} \times \mathbf{r}) = 2(h - U) \tag{18}$$

Substituting this result into Eq. (17) yields the following

$$J = \int_{\mathbf{r}_1}^{\mathbf{r}_2} d\rho,$$

from which it can be shown that the variation $\delta J$ along the orbit is zero. This result implies that the orbit on the manifold $H = h$ is the geodesic of $(\mathbf{A}^3, d\rho^2)$ with the metric $d\rho^2$, and the geodesic of $(\mathbf{A}^3, d\rho^2)$ with the metric $d\rho^2$ is the orbit on the manifold $H = h$. □

Denote $M = (\mathbf{A}^3, d\rho^2)$, where $M$ is a smooth manifold, and the metric $d\rho^2$ is not positive definite. For the equilibrium point $L \in M$, denote its tangent space as $T_L M$. Then, $\dim M = \dim T_L M = 3$. Let

$$TM = \bigcup_{p \in M} T_p M = \{(p, q) | p \in M, q \in T_p M\};$$

then, $TM$ is its tangent bundle. It follows that $\dim TM = 6$. Let $\Xi$ be a sufficiently small open neighborhood of the equilibrium point on the smooth manifold $M$. Then, the tangent bundle of $\Xi$ is $T\Xi = \bigcup_{p \in \Xi} T_p \Xi = \{(p, q) | p \in \Xi, q \in T_p \Xi\}$, and $\dim T\Xi = 6$. Let $(\mathbf{S}, \Omega)$ be a 6-dimensional symplectic manifold near the

equilibrium point such that $\mathbf{S}$ and $T\Xi$ are topological homeomorphism but not diffeomorphism, where $\Omega$ is a non-degenerate skew-symmetric bilinear quadratic form.

## 5. Manifold, Periodic Orbits and Quasi-periodic Orbits near Equilibrium Points

To determine the motion, the manifold, the periodic orbits and the quasi-periodic orbits near the equilibrium points, the stability and the eigenvalues of the equilibrium points must be known.

### 5.1 Eigenvalues and Submanifold

The equilibrium point in the potential field of a rotating asteroid has 6 eigenvalues, which are in the form of $\pm \alpha_j (\alpha \in \mathrm{R}, \alpha > 0; j = 1, 2, 3)$, $\pm i\beta_j (\beta \in \mathrm{R}, \beta > 0; j = 1, 2)$, and $\pm \sigma \pm i\tau (\sigma, \tau \in \mathrm{R}; \sigma, \tau > 0)$. In fact, the forms of the eigenvalues determine the structure of the submanifold and the subspace. There is a bijection between the form of the eigenvalues and the submanifold or the subspace.

Let us denote the Jacobian constant at the equilibrium point by $H(L)$, and the eigenvector of the eigenvalue $\lambda_j$ as $\mathbf{u}_j$.

Let us define the asymptotically stable, the asymptotically unstable and the centre manifold of the orbit on the manifold $H = h$ near the equilibrium point, where $h = H(L) + \varepsilon^2$, and $\varepsilon^2$ is sufficiently small that there is no other equilibrium point $\tilde{L}$ in the sufficiently small open neighborhood on the manifold $(\mathbf{S}, \Omega)$ with the Jacobian constant $H(\tilde{L})$ that satisfies $H(L) \leq H(\tilde{L}) \leq h$.

The asymptotically stable manifold $W^s(\mathbf{S})$, the asymptotically unstable manifold $W^u(\mathbf{S})$, and the centre manifold $W^c(\mathbf{S})$ are tangent to the asymptotically stable subspace $E^s(L)$, the asymptotically unstable subspace $E^u(L)$, and the centre subspace $E^c(L)$ at the equilibrium point, respectively, where

$$E^s(L) = span\{\mathbf{u}_j | \operatorname{Re}\lambda_j < 0\},$$

$$E^c(L) = span\{\mathbf{u}_j | \operatorname{Re}\lambda_j = 0\},$$

$$E^u(L) = span\{\mathbf{u}_j | \operatorname{Re}\lambda_j > 0\}.$$

Let's define $W^r(\mathbf{S})$ as the resonant manifold, which is tangent to the resonant subspace $E^r(L) = span\{\mathbf{u}_j | \exists \lambda_k, s.t. \operatorname{Re}\lambda_j = \operatorname{Re}\lambda_k = 0, \operatorname{Im}\lambda_j = \operatorname{Im}\lambda_k, j \neq k\}$.

Thus, it can be seen that $(\mathbf{S},\Omega) \simeq T\Xi \cong W^s(\mathbf{S}) \oplus W^c(\mathbf{S}) \oplus W^u(\mathbf{S})$, where $\simeq$ denotes a topological homeomorphism, $\cong$ denotes a diffeomorphism, and $\oplus$ denotes a direct sum. Then, $E^r(L) \subseteq E^c(L)$ and $W^r(\mathbf{S}) \subseteq W^c(\mathbf{S})$.

By defining $T_L\mathbf{S}$ as the tangent space of the manifold $(\mathbf{S},\Omega)$, the homeomorphism of the tangent space can be written as $T_L\mathbf{S} \cong E^s(L) \oplus E^c(L) \oplus E^u(L)$. Considering the dimension of the manifolds, the following equations hold

$$\dim W^s(\mathbf{S}) + \dim W^c(\mathbf{S}) + \dim W^u(\mathbf{S}) = \dim(\mathbf{S},\Omega) = \dim T\Xi = 6$$

$$\dim E^s(L) + \dim E^c(L) + \dim E^u(L) = \dim T_L\mathbf{S} = 6$$

$$\dim E^r(L) = \dim W^r(\mathbf{S}) \leq \dim E^c(L) = \dim W^c(\mathbf{S})$$

Based on the conclusions above, the following theorem can be obtained.

**Theorem 4.** There are eight cases for the equilibrium points in the potential field of a rotating asteroid:

**Case 1:** The eigenvalues are different and in the form of $\pm i\beta_j \left(\beta_j \in \mathrm{R}, \beta_j > 0; j = 1, 2, 3\right)$ ; then, the structure of the submanifold is $(\mathbf{S}, \Omega) \simeq T\Xi \cong W^c(\mathbf{S})$, and $\dim W^r(\mathbf{S}) = 0$.

**Case 2:** The forms of the eigenvalues are $\pm \alpha_j \left(\alpha_j \in \mathrm{R}, \alpha_j > 0, j = 1\right)$ and $\pm i\beta_j \left(\beta_j \in \mathrm{R}, \beta_j > 0; j = 1, 2\right)$, and the imaginary eigenvalues are different; then, the structure of the submanifold is $(\mathbf{S}, \Omega) \simeq T\Xi \cong W^s(\mathbf{S}) \oplus W^c(\mathbf{S}) \oplus W^u(\mathbf{S})$ , $\dim W^c(\mathbf{S}) = 4$, $\dim W^s(\mathbf{S}) = \dim W^u(\mathbf{S}) = 1$, and $\dim W^r(\mathbf{S}) = 0$.

**Case 3:** The forms of the eigenvalues are $\pm \alpha_j \left(\alpha_j \in \mathrm{R}, \alpha_j > 0; j = 1, 2\right)$ and $\pm i\beta_j \left(\beta_j \in \mathrm{R}, \beta_j > 0, j = 1\right)$ ; then, the structure of the submanifold is $(\mathbf{S}, \Omega) \simeq T\Xi \cong W^s(\mathbf{S}) \oplus W^c(\mathbf{S}) \oplus W^u(\mathbf{S})$ , $\dim W^r(\mathbf{S}) = 0$ , and $\dim W^s(\mathbf{S}) = \dim W^c(\mathbf{S}) = \dim W^u(\mathbf{S}) = 2$.

**Case 4:** The forms of the eigenvalues are $\pm \alpha_j \left(\alpha_j \in \mathrm{R}, \alpha_j > 0, j = 1\right)$ and $\pm \sigma \pm i\tau \left(\sigma, \tau \in \mathrm{R}; \sigma, \tau > 0\right)$ ; then, the structure of the submanifold is $(\mathbf{S}, \Omega) \simeq T\Xi \cong W^s(\mathbf{S}) \oplus W^u(\mathbf{S})$, and $\dim W^s(\mathbf{S}) = \dim W^u(\mathbf{S}) = 3$.

**Case 5:** The forms of the eigenvalues are $\pm i\beta_j \left(\beta_j \in \mathrm{R}, \beta_j > 0, j = 1\right)$ and $\pm \sigma \pm i\tau \left(\sigma, \tau \in \mathrm{R}; \sigma, \tau > 0\right)$ ; then, the structure of the submanifold is $(\mathbf{S}, \Omega) \simeq T\Xi \cong W^s(\mathbf{S}) \oplus W^c(\mathbf{S}) \oplus W^u(\mathbf{S})$ , $\dim W^r(\mathbf{S}) = 0$ , and $\dim W^s(\mathbf{S}) = \dim W^c(\mathbf{S}) = \dim W^u(\mathbf{S}) = 2$.

**Case 6:** The forms of the eigenvalues are $\pm i\beta_j \left(\beta_j \in \mathrm{R}, \beta_1 = \beta_2 = \beta_3 > 0; j = 1, 2, 3\right)$; then, the structure of the submanifold is $(\mathbf{S}, \Omega) \simeq T\Xi \simeq W^c(\mathbf{S}) \simeq W^r(\mathbf{S})$ , and $\dim W^r(\mathbf{S}) = \dim W^c(\mathbf{S}) = 6$.

**Case 7:** The forms of the eigenvalues are

$\pm i\beta_j \left( \beta_j \in \mathrm{R}, \beta_j > 0, \beta_1 = \beta_2 \neq \beta_3; j = 1, 2, 3 \right)$; then, the structure of the submanifold is $(\mathbf{S}, \Omega) \simeq T\Xi \cong W^c(\mathbf{S})$, and $\dim W^r(\mathbf{S}) = 4$.

**Case 8:** The forms of the eigenvalues are $\pm \alpha_j \left( \alpha_j \in \mathrm{R}, \alpha_j > 0, j = 1 \right), \pm i\beta_j \left( \beta_j \in \mathrm{R}, \beta_1 = \beta_2 > 0; j = 1, 2 \right)$; then, the structure of the submanifold is $(\mathbf{S}, \Omega) \simeq T\Xi \cong W^s(\mathbf{S}) \oplus W^c(\mathbf{S}) \oplus W^u(\mathbf{S})$, $\dim W^r(\mathbf{S}) = \dim W^c(\mathbf{S}) = 4$, and $\dim W^s(\mathbf{S}) = \dim W^u(\mathbf{S}) = 1$. □

Theorem 4 describes the structure of the submanifold as well as the stable and unstable behaviours of the particle near the equilibrium points. For Cases 6-8, because the resonant manifold and the resonant subspace exist, the equilibrium point is resonant. Considering that the structures of the submanifold and the subspace are fixed by the characteristic of the equilibrium points, it can be concluded that the equilibrium points with resonant manifolds are resonant equilibrium points. Only Case 1 leads to linearly stable equilibrium points, and Cases 2-5 lead to unstable equilibrium points. Thus, one can obtain

**Corollary 1.** The equilibrium point is linearly stable if and only if it belongs to Case 1. The equilibrium point is unstable and non-resonant if and only if it belongs to one of the Cases 2-5. The equilibrium point is resonant if and only if it belongs to one of the Cases 6-8.

The classes of orbit near the equilibrium points include: periodic orbit, Lissajous orbit, quasi-periodic orbit, almost periodic orbit, etc. Let $T^k$ be a $k$-dimensional torus. Then, the periodic orbit is on a 1-dimensional torus $T^1$, the Lissajous orbit is on a 2-dimensional torus $T^2$, and the quasi-periodic orbit is on a k-dimensional torus

$T^k \ (k \geq 1)$.

## 5.2 Linearly Stable Equilibrium Points

Theorem 4 and Corollary 1 show that the linearly stable equilibrium points only correspond to Case 1. In this section, more properties of Case 1 are discussed.

In Case 1, there are three pairs of imaginary eigenvalues of the equilibrium point, which is linearly stable. The motion of the spacecraft relative to the equilibrium point is a quasi-periodic orbit, which is expressed as

$$\begin{cases} \xi = C_{\xi 1} \cos \beta_1 t + S_{\xi 1} \sin \beta_1 t + C_{\xi 2} \cos \beta_2 t + S_{\xi 2} \sin \beta_2 t + C_{\xi 3} \cos \beta_3 t + S_{\xi 3} \sin \beta_3 t \\ \eta = C_{\eta 1} \cos \beta_1 t + S_{\eta 1} \sin \beta_1 t + C_{\eta 2} \cos \beta_2 t + S_{\eta 2} \sin \beta_2 t + C_{\eta 3} \cos \beta_3 t + S_{\eta 3} \sin \beta_3 t \\ \zeta = C_{\zeta 1} \cos \beta_1 t + S_{\zeta 1} \sin \beta_1 t + C_{\zeta 2} \cos \beta_2 t + S_{\zeta 2} \sin \beta_2 t + C_{\zeta 3} \cos \beta_3 t + S_{\zeta 3} \sin \beta_3 t \end{cases} \quad (19)$$

There are three families of periodic orbits, which have periods

$$T_1 = \frac{2\pi}{\beta_1}, T_2 = \frac{2\pi}{\beta_2}, T_3 = \frac{2\pi}{\beta_3} \tag{20}$$

With the condition

$C_{\xi 2} = C_{\eta 2} = C_{\zeta 2} = S_{\xi 2} = S_{\eta 2} = S_{\zeta 2} = C_{\xi 3} = C_{\eta 3} = C_{\zeta 3} = S_{\xi 3} = S_{\eta 3} = S_{\zeta 3} = 0$, the first family of periodic orbits has the form

$$\begin{cases} \xi = C_{\xi 1} \cos \beta_1 t + S_{\xi 1} \sin \beta_1 t \\ \eta = C_{\eta 1} \cos \beta_1 t + S_{\eta 1} \sin \beta_1 t \\ \zeta = C_{\zeta 1} \cos \beta_1 t + S_{\zeta 1} \sin \beta_1 t \end{cases} \tag{21}$$

The period of the first family of periodic orbits is $T_1 = \dfrac{2\pi}{\beta_1}$.

Denote $t_0$ as the initial time. Then, the initial state can be expressed as

$$\begin{cases} \xi(t_0) = \xi_0, \dot{\xi}(t_0) = \dot{\xi}_0 \\ \eta(t_0) = \eta_0, \dot{\eta}(t_0) = \dot{\eta}_0 \\ \zeta(t_0) = \zeta_0, \dot{\zeta}(t_0) = \dot{\zeta}_0 \end{cases} \tag{22}$$

and the coefficients yield

$$\begin{cases} C_{\xi 1} = \xi_0 \cos \beta_1 t_0 - \dfrac{\dot{\xi}_0}{\beta_1} \sin \beta_1 t \\ S_{\xi 1} = \xi_0 \sin \beta_1 t_0 + \dfrac{\dot{\xi}_0}{\beta_1} \cos \beta_1 t \\ C_{\eta 1} = \eta_0 \cos \beta_1 t_0 - \dfrac{\dot{\eta}_0}{\beta_1} \sin \beta_1 t \\ S_{\eta 1} = \eta_0 \sin \beta_1 t_0 + \dfrac{\dot{\eta}_0}{\beta_1} \cos \beta_1 t \\ C_{\zeta 1} = \zeta_0 \cos \beta_1 t_0 - \dfrac{\dot{\zeta}_0}{\beta_1} \sin \beta_1 t \\ S_{\zeta 1} = \zeta_0 \sin \beta_1 t_0 + \dfrac{\dot{\zeta}_0}{\beta_1} \cos \beta_1 t \end{cases} \quad (23)$$

Using Eq. (23), we can calculate the coefficients of the first families of periodic orbits if we know the initial state. The other two families of periodic orbits satisfy the condition

$C_{\xi 1} = C_{\eta 1} = C_{\zeta 1} = S_{\xi 1} = S_{\eta 1} = S_{\zeta 1} = C_{\xi 3} = C_{\eta 3} = C_{\zeta 3} = S_{\xi 3} = S_{\eta 3} = S_{\zeta 3} = 0$, or

$C_{\xi 1} = C_{\eta 1} = C_{\zeta 1} = S_{\xi 1} = S_{\eta 1} = S_{\zeta 1} = C_{\xi 2} = C_{\eta 2} = C_{\zeta 2} = S_{\xi 2} = S_{\eta 2} = S_{\zeta 2} = 0$

In addition, they have similar forms of the position equation and the coefficient equation.

**Theorem 5.** For an equilibrium point in the potential field of a rotating asteroid, the following conditions are equivalent:

a) It is linearly stable.

b) The roots of the characteristic equation $P(\lambda)$ are in the form of $\pm i\beta_j \left( \beta_j \in \mathrm{R}, \beta_j > 0; j = 1, 2, 3 \right)$, and if $j \neq k (j = 1, 2, 3; k = 1, 2, 3)$, then $\beta_j \neq \beta_k$.

c) The motion of the spacecraft relative to the equilibrium point is a quasi-periodic

orbit, which is expressed as

$$\begin{cases} \xi = C_{\xi 1}\cos\beta_1 t + S_{\xi 1}\sin\beta_1 t + C_{\xi 2}\cos\beta_2 t + S_{\xi 2}\sin\beta_2 t + C_{\xi 3}\cos\beta_3 t + S_{\xi 3}\sin\beta_3 t \\ \eta = C_{\eta 1}\cos\beta_1 t + S_{\eta 1}\sin\beta_1 t + C_{\eta 2}\cos\beta_2 t + S_{\eta 2}\sin\beta_2 t + C_{\eta 3}\cos\beta_3 t + S_{\eta 3}\sin\beta_3 t \\ \zeta = C_{\zeta 1}\cos\beta_1 t + S_{\zeta 1}\sin\beta_1 t + C_{\zeta 2}\cos\beta_2 t + S_{\zeta 2}\sin\beta_2 t + C_{\zeta 3}\cos\beta_3 t + S_{\zeta 3}\sin\beta_3 t \end{cases}$$

d) There are four families of quasi-periodic orbits in the tangent space of the equilibrium point, which can be expressed as

$$\begin{cases} \xi = C_{\xi 1}\cos\beta_1 t + S_{\xi 1}\sin\beta_1 t + C_{\xi 2}\cos\beta_2 t + S_{\xi 2}\sin\beta_2 t \\ \eta = C_{\eta 1}\cos\beta_1 t + S_{\eta 1}\sin\beta_1 t + C_{\eta 2}\cos\beta_2 t + S_{\eta 2}\sin\beta_2 t \\ \zeta = C_{\zeta 1}\cos\beta_1 t + S_{\zeta 1}\sin\beta_1 t + C_{\zeta 2}\cos\beta_2 t + S_{\zeta 2}\sin\beta_2 t \end{cases}$$

$$\begin{cases} \xi = C_{\xi 1}\cos\beta_1 t + S_{\xi 1}\sin\beta_1 t + C_{\xi 3}\cos\beta_3 t + S_{\xi 3}\sin\beta_3 t \\ \eta = C_{\eta 1}\cos\beta_1 t + S_{\eta 1}\sin\beta_1 t + C_{\eta 3}\cos\beta_3 t + S_{\eta 3}\sin\beta_3 t \\ \zeta = C_{\zeta 1}\cos\beta_1 t + S_{\zeta 1}\sin\beta_1 t + C_{\zeta 3}\cos\beta_3 t + S_{\zeta 3}\sin\beta_3 t \end{cases}$$

$$\begin{cases} \xi = C_{\xi 2}\cos\beta_2 t + S_{\xi 2}\sin\beta_2 t + C_{\xi 3}\cos\beta_3 t + S_{\xi 3}\sin\beta_3 t \\ \eta = C_{\eta 2}\cos\beta_2 t + S_{\eta 2}\sin\beta_2 t + C_{\eta 3}\cos\beta_3 t + S_{\eta 3}\sin\beta_3 t \\ \zeta = C_{\zeta 2}\cos\beta_2 t + S_{\zeta 2}\sin\beta_2 t + C_{\zeta 3}\cos\beta_3 t + S_{\zeta 3}\sin\beta_3 t \end{cases}$$

$$\begin{cases} \xi = C_{\xi 1}\cos\beta_1 t + S_{\xi 1}\sin\beta_1 t + C_{\xi 2}\cos\beta_2 t + S_{\xi 2}\sin\beta_2 t + C_{\xi 3}\cos\beta_3 t + S_{\xi 3}\sin\beta_3 t \\ \eta = C_{\eta 1}\cos\beta_1 t + S_{\eta 1}\sin\beta_1 t + C_{\eta 2}\cos\beta_2 t + S_{\eta 2}\sin\beta_2 t + C_{\eta 3}\cos\beta_3 t + S_{\eta 3}\sin\beta_3 t \\ \zeta = C_{\zeta 1}\cos\beta_1 t + S_{\zeta 1}\sin\beta_1 t + C_{\zeta 2}\cos\beta_2 t + S_{\zeta 2}\sin\beta_2 t + C_{\zeta 3}\cos\beta_3 t + S_{\zeta 3}\sin\beta_3 t \end{cases}$$

e) There are four families of quasi-periodic orbits near the equilibrium point, and they are on the k-dimensional tori $T^k (k=2,3)$.

f) There are three families of periodic orbits in the tangent space of the equilibrium point, which have the periods $T_1 = \frac{2\pi}{\beta_1}, T_2 = \frac{2\pi}{\beta_2}, T_3 = \frac{2\pi}{\beta_3}$.

g) There are three families of periodic orbits near the equilibrium point.

h) There is no asymptotically stable manifold near the equilibrium point, and the roots of the characteristic equation $P(\lambda)$ satisfy the following condition: If $j \neq k (j=1,2,\cdots,6; k=1,2,\cdots,6)$, then $\lambda_j \neq \lambda_k$.

i) There is no unstable manifold near the equilibrium point, and the roots of the

characteristic equation $P(\lambda)$ satisfy the following condition: If $j \neq k\,(j=1,2,\cdots,6; k=1,2,\cdots,6)$, then $\lambda_j \neq \lambda_k$.

j) The dimensions of the centre manifold $W^c(\mathbf{S})$ and the resonant manifold $W^r(\mathbf{S})$ satisfy $\dim W^c(\mathbf{S}) = 6$ and $\dim W^r(\mathbf{S}) = 0$, respectively.

k) The dimensions of the centre subspace $E^c(L)$ and the resonant subspace $E^r(L)$ satisfy $\dim E^c(L) = 6$ and $\dim E^r(L) = 0$, respectively.

l) The structure of the submanifold is $(\mathbf{S}, \Omega) \simeq T\Xi \cong W^c(\mathbf{S})$, $W^r(\mathbf{S}) = \varnothing$, where $\varnothing$ is an empty set.

m) The structure of the subspace is $T_L \mathbf{S} \cong E^c(L)$, and $E^r(L) = \varnothing$.

**Proof:**

a)$\Rightarrow$b): Using Theorem 4, it is obvious.

b)$\Rightarrow$c): Using the solution theory of ordinary differential equations, one can obtain c).

c)$\Rightarrow$d): It is obvious.

d)$\Rightarrow$e): Among the four families of quasi-periodic orbits that are expressed in d), the first three families of quasi-periodic orbits near the equilibrium point are on 2-dimensional tori $T^2$, and the last family of quasi-periodic orbits near the equilibrium point is on a 3-dimensional torus $T^3$.

e)$\Rightarrow$f): Assuming that there are four families of quasi-periodic orbits near the equilibrium point, which are on the k-dimensional tori $T^k\,(k=2,3)$, we can eliminate Cases 2-8. Thus, the equilibrium point is linearly stable, and a) is established. Considering a) and c), we derive that there are three families of periodic orbits in the tangent space of the equilibrium point that have the periods $T_1 = \frac{2\pi}{\beta_1}, T_2 = \frac{2\pi}{\beta_2}$ and $T_3 = \frac{2\pi}{\beta_3}$.

f)$\Rightarrow$g): It is obvious.

g)$\Rightarrow$a): Because there are three families of periodic orbits, we can eliminate Cases 2-8 and obtain a).

a)$\Rightarrow$i): Because the equilibrium point is linearly stable, there is no unstable manifold or resonant manifold near the equilibrium point. We obtain i).

h)$\Leftrightarrow$i)$\Leftrightarrow$j)$\Leftrightarrow$k)$\Leftrightarrow$l)$\Leftrightarrow$m): It is obvious.

m)$\Rightarrow$a): The structure of the subspace is $T_L\mathbf{S} \cong E^c(L)$, and $E^r(L) = \varnothing$. Thus, the dimensions of the unstable manifold and the resonant manifold near the equilibrium point are zero. Then, the eigenvalues have the form $\pm i\beta_j \left(\beta_j \in \mathbb{R}, \beta_j > 0; j=1,2,3\right)$, and if $j \neq k (j=1,2,3; k=1,2,3)$, then $\beta_j \neq \beta_k$. This result leads to a). □

Strictly speaking, Eq. (19) is the projection of the motion to the tangent space of the equilibrium point; Eq. (19) is the approximate expression of the motion. Because Theorem 6 is about the necessary and sufficient conditions of the linear stability of the equilibrium points in the potential field of a rotating asteroid, there is a corollary for the linear instability of the equilibrium points.

**Corollary 2.** For an equilibrium point in the potential field of a rotating asteroid, the following conditions are equivalent:

a) It is linearly unstable.

b) The roots of the characteristic equation $P(\lambda)$ do not have the form

$\pm i\beta_j \left(\beta_j \in \mathrm{R}, \beta_j > 0; \beta_j \neq \beta_k; j, k = 1, 2, 3, j \neq k \right)$.

c) There are fewer than four families of quasi-periodic orbits on the k-dimensional tori $T^k (k=2,3)$ near the equilibrium point.

d) There are fewer than three families of periodic orbits in the tangent space of the equilibrium point.

e) There is an asymptotically stable manifold near the equilibrium point or the roots of the characteristic equation $P(\lambda)$ that satisfies the following condition: If $j \neq k (j = 1, 2, \cdots, 6; k = 1, 2, \cdots, 6)$, then $\lambda_j \neq \lambda_k$.

f) There is an unstable manifold near the equilibrium point or the roots of the characteristic equation $P(\lambda)$ that satisfies the following condition: If $j \neq k (j = 1, 2, \cdots, 6; k = 1, 2, \cdots, 6)$, then $\lambda_j \neq \lambda_k$.

g) $\dim W^c(\mathbf{S}) < 6$ or $\begin{cases} \dim W^c(\mathbf{S}) = 6 \\ \dim W^r(\mathbf{S}) > 0 \end{cases}$.

h) $\dim E^c(L) < 6$ or $\begin{cases} \dim E^c(L) = 6 \\ \dim E^r(L) > 0 \end{cases}$. $\square$

## 5.3 Unstable Equilibrium Points

In this section, the unstable equilibrium points with no resonant manifold are discussed. The unstable equilibrium points can be classified into four cases, which have been presented in Theorem 4. From Theorem 4, it is known that an equilibrium point is unstable if and only if $\dim W^u(\mathbf{S}) \geq 1, \dim W^r(\mathbf{S}) = 0$, with

Case 2 corresponding to $\dim W^u(\mathbf{S}) = 1$ and $\dim W^r(\mathbf{S}) = 0$;

Case 3 and Case 5 corresponding to $\dim W^u(\mathbf{S}) = 2$ and $\dim W^r(\mathbf{S}) = 0$;

Case 4 corresponding to $\dim W^u(\mathbf{S}) = 3$ and $\dim W^r(\mathbf{S}) = 0$.

**5.3.1 Case 2**

There are two pairs of imaginary eigenvalues and one pair of real eigenvalues for the unstable equilibrium point. The motion of the spacecraft near this equilibrium point relative to this equilibrium point is expressed by

$$\begin{cases} \xi = A_{\xi 1} e^{\alpha_1 t} + B_{\xi 1} e^{-\alpha_1 t} + C_{\xi 1} \cos \beta_1 t + S_{\xi 1} \sin \beta_1 t + C_{\xi 2} \cos \beta_2 t + S_{\xi 2} \sin \beta_2 t \\ \eta = A_{\eta 1} e^{\alpha_1 t} + B_{\eta 1} e^{-\alpha_1 t} + C_{\eta 1} \cos \beta_1 t + S_{\eta 1} \sin \beta_1 t + C_{\eta 2} \cos \beta_2 t + S_{\eta 2} \sin \beta_2 t \\ \zeta = A_{\zeta 1} e^{\alpha_1 t} + B_{\zeta 1} e^{-\alpha_1 t} + C_{\zeta 1} \cos \beta_1 t + S_{\zeta 1} \sin \beta_1 t + C_{\zeta 2} \cos \beta_2 t + S_{\zeta 2} \sin \beta_2 t \end{cases} \quad (24)$$

The almost periodic orbit near the equilibrium point can be expressed by

$$\begin{cases} \xi = B_{\xi 1} e^{-\alpha_1 t} + C_{\xi 1} \cos \beta_1 t + S_{\xi 1} \sin \beta_1 t + C_{\xi 2} \cos \beta_2 t + S_{\xi 2} \sin \beta_2 t \\ \eta = B_{\eta 1} e^{-\alpha_1 t} + C_{\eta 1} \cos \beta_1 t + S_{\eta 1} \sin \beta_1 t + C_{\eta 2} \cos \beta_2 t + S_{\eta 2} \sin \beta_2 t \\ \zeta = B_{\zeta 1} e^{-\alpha_1 t} + C_{\zeta 1} \cos \beta_1 t + S_{\zeta 1} \sin \beta_1 t + C_{\zeta 2} \cos \beta_2 t + S_{\zeta 2} \sin \beta_2 t \end{cases} \quad (25)$$

The central manifold is a 4-dimensional smooth manifold, which is given by $A_{\xi 1} = B_{\xi 1} = A_{\eta 1} = B_{\eta 1} = A_{\zeta 1} = B_{\zeta 1} = 0$. The motion of the spacecraft relative to the equilibrium point on the central manifold is a Lissajous orbit.

There are two families of periodic orbits on the central manifold. The first family of periodic orbits is given by $\begin{cases} A_{\xi 1} = B_{\xi 1} = A_{\eta 1} = B_{\eta 1} = A_{\zeta 1} = B_{\zeta 1} = 0 \\ C_{\xi 2} = S_{\xi 2} = C_{\eta 2} = S_{\eta 2} = C_{\zeta 2} = S_{\zeta 2} = 0 \end{cases}$, and the period is $T_1 = \dfrac{2\pi}{\beta_1}$. The second family of periodic orbits is given by $\begin{cases} A_{\xi 1} = B_{\xi 1} = A_{\eta 1} = B_{\eta 1} = A_{\zeta 1} = B_{\zeta 1} = 0 \\ C_{\xi 1} = S_{\xi 1} = C_{\eta 1} = S_{\eta 1} = C_{\zeta 1} = S_{\zeta 1} = 0 \end{cases}$, and the period is $T_2 = \dfrac{2\pi}{\beta_2}$.

The asymptotically stable manifold is generated by

$$\begin{cases} \xi = B_{\xi 1} e^{-\alpha_1 t} \\ \eta = B_{\eta 1} e^{-\alpha_1 t} \\ \zeta = B_{\zeta 1} e^{-\alpha_1 t} \end{cases} \quad (26)$$

It is a 1-dimensional smooth manifold.

The unstable manifold is generated by

$$\begin{cases} \xi = A_{\xi 1} e^{\alpha_1 t} \\ \eta = A_{\eta 1} e^{\alpha_1 t} \\ \zeta = A_{\zeta 1} e^{\alpha_1 t} \end{cases} \quad (27)$$

It is a 1-dimensional smooth manifold. The general result of Case 2 is stated as follows.

**Theorem 6.** For an equilibrium point in the potential field of a rotating asteroid, the following conditions are equivalent:

a) The roots of the characteristic equation $P(\lambda)$ are in the form of $\pm \alpha_j \left( \alpha_j \in \mathrm{R}, \alpha_j > 0, j = 1 \right)$ and $\pm i \beta_j \left( \beta_j \in \mathrm{R}, \beta_j > 0; j = 1, 2 \right)$, where $\beta_1 \neq \beta_2$.

b) The motion of the spacecraft near the equilibrium point relative to the equilibrium point can be expressed as

$$\begin{cases} \xi = A_{\xi 1} e^{\alpha_1 t} + B_{\xi 1} e^{-\alpha_1 t} + C_{\xi 1} \cos \beta_1 t + S_{\xi 1} \sin \beta_1 t + C_{\xi 2} \cos \beta_2 t + S_{\xi 2} \sin \beta_2 t \\ \eta = A_{\eta 1} e^{\alpha_1 t} + B_{\eta 1} e^{-\alpha_1 t} + C_{\eta 1} \cos \beta_1 t + S_{\eta 1} \sin \beta_1 t + C_{\eta 2} \cos \beta_2 t + S_{\eta 2} \sin \beta_2 t \\ \zeta = A_{\zeta 1} e^{\alpha_1 t} + B_{\zeta 1} e^{-\alpha_1 t} + C_{\zeta 1} \cos \beta_1 t + S_{\zeta 1} \sin \beta_1 t + C_{\zeta 2} \cos \beta_2 t + S_{\zeta 2} \sin \beta_2 t \end{cases}$$

c) The characteristic roots are different, and there are two families of quasi-periodic orbits in the tangent space of the equilibrium point, which can be expressed as

$$\begin{cases} \xi = C_{\xi 1} \cos \beta_1 t + S_{\xi 1} \sin \beta_1 t + C_{\xi 2} \cos \beta_2 t + S_{\xi 2} \sin \beta_2 t \\ \eta = C_{\eta 1} \cos \beta_1 t + S_{\eta 1} \sin \beta_1 t + C_{\eta 2} \cos \beta_2 t + S_{\eta 2} \sin \beta_2 t \\ \zeta = C_{\zeta 1} \cos \beta_1 t + S_{\zeta 1} \sin \beta_1 t + C_{\zeta 2} \cos \beta_2 t + S_{\zeta 2} \sin \beta_2 t \end{cases}$$

d) The characteristic roots are different, and there are two families of quasi-periodic

orbits near the equilibrium point on the 2-dimensional tori $T^2$.

e) There are two families of periodic orbits in the tangent space of the equilibrium point, which have the periods $T_1 = \dfrac{2\pi}{\beta_1}, T_2 = \dfrac{2\pi}{\beta_2}$.

f) There are two families of periodic orbits near the equilibrium point.

g) The structure of the submanifold is $(\mathbf{S},\Omega) \simeq T\Xi \cong W^s(\mathbf{S}) \oplus W^c(\mathbf{S}) \oplus W^u(\mathbf{S})$; $\dim W^c(\mathbf{S}) = 4$, $\dim W^s(\mathbf{S}) = \dim W^u(\mathbf{S}) = 1$, and $\dim W^r(\mathbf{S}) = 0$.

h) The structure of the subspace is $T_L\mathbf{S} \cong E^s(L) \oplus E^c(L) \oplus E^u(L)$; $\dim E^c(L) = 4$, $\dim E^s(L) = \dim E^u(L) = 1$, and $\dim W^r(\mathbf{S}) = 0$.

**Proof:**

The proof for Theorem 6 is similar to that for Theorem 5. □

### 5.3.2 Case 3

The eigenvalues of the equilibrium point are in the form of $\pm\alpha_j \left(\alpha_j \in \mathrm{R}, \alpha_j > 0; j = 1, 2\right)$ and $\pm i\beta_j \left(\beta_j \in \mathrm{R}, \beta_j > 0, j = 1\right)$. The equilibrium point is unstable. The motion of the spacecraft near the equilibrium point relative to the equilibrium point is expressed as

$$\begin{cases} \xi = A_{\xi 1}e^{\alpha_1 t} + B_{\xi 1}e^{-\alpha_1 t} + A_{\xi 2}e^{\alpha_2 t} + B_{\xi 2}e^{-\alpha_2 t} + C_{\xi 1}\cos\beta_1 t + S_{\xi 1}\sin\beta_1 t \\ \eta = A_{\eta 1}e^{\alpha_1 t} + B_{\eta 1}e^{-\alpha_1 t} + A_{\eta 2}e^{\alpha_2 t} + B_{\eta 2}e^{-\alpha_2 t} + C_{\eta 1}\cos\beta_1 t + S_{\eta 1}\sin\beta_1 t \\ \zeta = A_{\zeta 1}e^{\alpha_1 t} + B_{\zeta 1}e^{-\alpha_1 t} + A_{\zeta 2}e^{\alpha_2 t} + B_{\zeta 2}e^{-\alpha_2 t} + C_{\zeta 1}\cos\beta_1 t + S_{\zeta 1}\sin\beta_1 t \end{cases} \quad (28)$$

The almost periodic orbit near the equilibrium point can be expressed as

$$\begin{cases} \xi = B_{\xi 1}e^{-\alpha_1 t} + B_{\xi 2}e^{-\alpha_2 t} + C_{\xi 1}\cos\beta_1 t + S_{\xi 1}\sin\beta_1 t \\ \eta = B_{\eta 1}e^{-\alpha_1 t} + B_{\eta 2}e^{-\alpha_2 t} + C_{\eta 1}\cos\beta_1 t + S_{\eta 1}\sin\beta_1 t \\ \zeta = B_{\zeta 1}e^{-\alpha_1 t} + B_{\zeta 2}e^{-\alpha_2 t} + C_{\zeta 1}\cos\beta_1 t + S_{\zeta 1}\sin\beta_1 t \end{cases} \quad (29)$$

The central manifold is given by $\begin{cases} A_{\xi 1} = B_{\xi 1} = A_{\eta 1} = B_{\eta 1} = A_{\zeta 1} = B_{\zeta 1} = 0 \\ A_{\xi 2} = B_{\xi 2} = A_{\eta 2} = B_{\eta 2} = A_{\zeta 2} = B_{\zeta 2} = 0 \end{cases}$, and it is a 2-dimensional smooth manifold.

There is one family of periodic orbits, which is given by

$\begin{cases} A_{\xi 1} = B_{\xi 1} = A_{\eta 1} = B_{\eta 1} = A_{\zeta 1} = B_{\zeta 1} = 0 \\ A_{\xi 2} = B_{\xi 2} = A_{\eta 2} = B_{\eta 2} = A_{\zeta 2} = B_{\zeta 2} = 0 \end{cases}$, and the period is $T_1 = \dfrac{2\pi}{\beta_1}$.

The asymptotically stable manifold is generated by

$$\begin{cases} \xi = B_{\xi 1} e^{-\alpha_1 t} + B_{\xi 2} e^{-\alpha_2 t} \\ \eta = B_{\eta 1} e^{-\alpha_1 t} + B_{\eta 2} e^{-\alpha_2 t} \\ \zeta = B_{\zeta 1} e^{-\alpha_1 t} + B_{\zeta 2} e^{-\alpha_2 t} \end{cases} \quad (30)$$

It is a 2-dimensional smooth manifold.

The unstable manifold is generated by

$$\begin{cases} \xi = A_{\xi 1} e^{\alpha_1 t} + A_{\xi 2} e^{\alpha_2 t} \\ \eta = A_{\eta 1} e^{\alpha_1 t} + A_{\eta 2} e^{\alpha_2 t} \\ \zeta = A_{\zeta 1} e^{\alpha_1 t} + A_{\zeta 2} e^{\alpha_2 t} \end{cases} \quad (31)$$

It is a 2-dimensional smooth manifold. Then, the general result for Case 3 is stated as follows.

**Theorem 7.** For an equilibrium point in the potential field of a rotating asteroid, the following conditions are equivalent:

a) The roots of the characteristic equation $P(\lambda)$ have the form $\pm \alpha_j \left( \alpha_j \in \mathrm{R}, \alpha_j > 0; j = 1, 2 \right)$ and $\pm i\beta_j \left( \beta_j \in \mathrm{R}, \beta_j > 0, j = 1 \right)$.

b) The motion of the spacecraft near the equilibrium point relative to the equilibrium point can be expressed as

$$\begin{cases} \xi = A_{\xi 1}e^{\alpha_1 t} + B_{\xi 1}e^{-\alpha_1 t} + A_{\xi 2}e^{\alpha_2 t} + B_{\xi 2}e^{-\alpha_2 t} + C_{\xi 1}\cos\beta_1 t + S_{\xi 1}\sin\beta_1 t \\ \eta = A_{\eta 1}e^{\alpha_1 t} + B_{\eta 1}e^{-\alpha_1 t} + A_{\eta 2}e^{\alpha_2 t} + B_{\eta 2}e^{-\alpha_2 t} + C_{\eta 1}\cos\beta_1 t + S_{\eta 1}\sin\beta_1 t \\ \zeta = A_{\zeta 1}e^{\alpha_1 t} + B_{\zeta 1}e^{-\alpha_1 t} + A_{\zeta 2}e^{\alpha_2 t} + B_{\zeta 2}e^{-\alpha_2 t} + C_{\zeta 1}\cos\beta_1 t + S_{\zeta 1}\sin\beta_1 t \end{cases}$$

c) There is one family of periodic orbits in the tangent space of the equilibrium point, which has the period $T_1 = \dfrac{2\pi}{\beta_1}$, and the dimension of the unstable manifold satisfies $\dim W^u(\mathbf{S}) = 2$; there is at least one characteristic root in the real axis.

d) There is one family of periodic orbits near the equilibrium point, and the dimension of the unstable manifold satisfies $\dim W^u(\mathbf{S}) = 2$; there is at least one characteristic root in the real axis.

e) The asymptotically stable manifold is generated by

$$\begin{cases} \xi = B_{\xi 1}e^{-\alpha_1 t} + B_{\xi 2}e^{-\alpha_2 t} \\ \eta = B_{\eta 1}e^{-\alpha_1 t} + B_{\eta 2}e^{-\alpha_2 t} \\ \zeta = B_{\zeta 1}e^{-\alpha_1 t} + B_{\zeta 2}e^{-\alpha_2 t} \end{cases}.$$

f) The unstable manifold is generated by

$$\begin{cases} \xi = A_{\xi 1}e^{\alpha_1 t} + A_{\xi 2}e^{\alpha_2 t} \\ \eta = A_{\eta 1}e^{\alpha_1 t} + A_{\eta 2}e^{\alpha_2 t} \\ \zeta = A_{\zeta 1}e^{\alpha_1 t} + A_{\zeta 2}e^{\alpha_2 t} \end{cases}.$$

g) The structure of the submanifold is $(\mathbf{S}, \Omega) \simeq T\Xi \cong W^s(\mathbf{S}) \oplus W^c(\mathbf{S}) \oplus W^u(\mathbf{S})$, and $\dim W^s(\mathbf{S}) = \dim W^c(\mathbf{S}) = \dim W^u(\mathbf{S}) = 2$; there is at least one characteristic root in the real axis.

h) The structure of the subspace is $T_L\mathbf{S} \cong E^s(L) \oplus E^c(L) \oplus E^u(L)$, and $\dim E^s(L) = \dim E^c(L) = \dim E^u(L) = 2$; there is at least one characteristic root in the real axis. □

## 5.3.3 Case 4

The forms of the eigenvalues are $\pm\alpha_j\left(\alpha_j \in \mathrm{R}, \alpha_j > 0, j=1\right)$ and $\pm\sigma \pm i\tau\left(\sigma, \tau \in \mathrm{R}; \sigma, \tau > 0\right)$. The equilibrium point is unstable. The motion of the spacecraft near the equilibrium point relative to the equilibrium point is expressed as

$$\begin{cases} \xi = A_{\xi 1}e^{\alpha_1 t} + B_{\xi 1}e^{-\alpha_1 t} + E_\xi e^{\sigma_1 t}\cos\tau_1 t + F_\xi e^{\sigma_1 t}\sin\tau_1 t + G_\xi e^{-\sigma_1 t}\cos\tau_1 t + H_\xi e^{-\sigma_1 t}\sin\tau_1 t \\ \eta = A_{\eta 1}e^{\alpha_1 t} + B_{\eta 1}e^{-\alpha_1 t} + E_\eta e^{\sigma_1 t}\cos\tau_1 t + F_\eta e^{\sigma_1 t}\sin\tau_1 t + G_\eta e^{-\sigma_1 t}\cos\tau_1 t + H_\eta e^{-\sigma_1 t}\sin\tau_1 t \\ \zeta = A_{\zeta 1}e^{\alpha_1 t} + B_{\zeta 1}e^{-\alpha_1 t} + E_\zeta e^{\sigma_1 t}\cos\tau_1 t + F_\zeta e^{\sigma_1 t}\sin\tau_1 t + G_\zeta e^{-\sigma_1 t}\cos\tau_1 t + H_\zeta e^{-\sigma_1 t}\sin\tau_1 t \end{cases} \quad (32)$$

The asymptotically stable manifold is generated by

$$\begin{cases} \xi = B_{\xi 1}e^{-\alpha_1 t} + G_\xi e^{-\sigma_1 t}\cos\tau_1 t + H_\xi e^{-\sigma_1 t}\sin\tau_1 t \\ \eta = B_{\eta 1}e^{-\alpha_1 t} + G_\eta e^{-\sigma_1 t}\cos\tau_1 t + H_\eta e^{-\sigma_1 t}\sin\tau_1 t \\ \zeta = B_{\zeta 1}e^{-\alpha_1 t} + G_\zeta e^{-\sigma_1 t}\cos\tau_1 t + H_\zeta e^{-\sigma_1 t}\sin\tau_1 t \end{cases} \quad (33)$$

It is a 3-dimensional smooth manifold. The unstable manifold is generated by

$$\begin{cases} \xi = A_{\xi 1}e^{\alpha_1 t} + E_\xi e^{\sigma_1 t}\cos\tau_1 t + F_\xi e^{\sigma_1 t}\sin\tau_1 t \\ \eta = A_{\eta 1}e^{\alpha_1 t} + E_\eta e^{\sigma_1 t}\cos\tau_1 t + F_\eta e^{\sigma_1 t}\sin\tau_1 t \\ \zeta = A_{\zeta 1}e^{\alpha_1 t} + E_\zeta e^{\sigma_1 t}\cos\tau_1 t + F_\zeta e^{\sigma_1 t}\sin\tau_1 t \end{cases} \quad (34)$$

It is a 3-dimensional smooth manifold. Then, the general result of Case 4 is stated as follows.

**Theorem 8.** For an equilibrium point in the potential field of a rotating asteroid, the following conditions are equivalent:

a) The roots of the characteristic equation $P(\lambda)$ are in the form of $\pm\alpha_j\left(\alpha_j \in \mathrm{R}, \alpha_j > 0, j=1\right)$ and $\pm\sigma \pm i\tau\left(\sigma, \tau \in \mathrm{R}; \sigma, \tau > 0\right)$.

b) The motion of the spacecraft near the equilibrium point relative to the equilibrium point can be expressed as

$$\begin{cases} \xi = A_{\xi 1}e^{\alpha_1 t} + B_{\xi 1}e^{-\alpha_1 t} + E_\xi e^{\sigma_1 t}\cos\tau_1 t + F_\xi e^{\sigma_1 t}\sin\tau_1 t + G_\xi e^{-\sigma_1 t}\cos\tau_1 t + H_\xi e^{-\sigma_1 t}\sin\tau_1 t \\ \eta = A_{\eta 1}e^{\alpha_1 t} + B_{\eta 1}e^{-\alpha_1 t} + E_\eta e^{\sigma_1 t}\cos\tau_1 t + F_\eta e^{\sigma_1 t}\sin\tau_1 t + G_\eta e^{-\sigma_1 t}\cos\tau_1 t + H_\eta e^{-\sigma_1 t}\sin\tau_1 t \\ \zeta = A_{\zeta 1}e^{\alpha_1 t} + B_{\zeta 1}e^{-\alpha_1 t} + E_\zeta e^{\sigma_1 t}\cos\tau_1 t + F_\zeta e^{\sigma_1 t}\sin\tau_1 t + G_\zeta e^{-\sigma_1 t}\cos\tau_1 t + H_\zeta e^{-\sigma_1 t}\sin\tau_1 t \end{cases}$$

c) There is no periodic orbit in the tangent space of the equilibrium point.

d) There is no periodic orbit near the equilibrium point.

e) The asymptotically stable manifold is generated by

$$\begin{cases} \xi = B_{\xi 1}e^{-\alpha_1 t} + G_\xi e^{-\sigma_1 t}\cos\tau_1 t + H_\xi e^{-\sigma_1 t}\sin\tau_1 t \\ \eta = B_{\eta 1}e^{-\alpha_1 t} + G_\eta e^{-\sigma_1 t}\cos\tau_1 t + H_\eta e^{-\sigma_1 t}\sin\tau_1 t \\ \zeta = B_{\zeta 1}e^{-\alpha_1 t} + G_\zeta e^{-\sigma_1 t}\cos\tau_1 t + H_\zeta e^{-\sigma_1 t}\sin\tau_1 t \end{cases}.$$

f) The unstable manifold is generated by

$$\begin{cases} \xi = A_{\xi 1}e^{\alpha_1 t} + E_\xi e^{\sigma_1 t}\cos\tau_1 t + F_\xi e^{\sigma_1 t}\sin\tau_1 t \\ \eta = A_{\eta 1}e^{\alpha_1 t} + E_\eta e^{\sigma_1 t}\cos\tau_1 t + F_\eta e^{\sigma_1 t}\sin\tau_1 t \\ \zeta = A_{\zeta 1}e^{\alpha_1 t} + E_\zeta e^{\sigma_1 t}\cos\tau_1 t + F_\zeta e^{\sigma_1 t}\sin\tau_1 t \end{cases}.$$

g) The structure of the submanifold is $(\mathbf{S},\Omega) \simeq T\Xi \cong W^s(\mathbf{S}) \oplus W^u(\mathbf{S})$.

h) The dimension of the unstable manifold satisfies $\dim W^u(\mathbf{S}) = 3$.

i) The dimension of the asymptotically stable manifold satisfies $\dim W^s(\mathbf{S}) = 3$

j) The structure of the subspace is $T_L \mathbf{S} \cong E^s(L) \oplus E^u(L)$.

k) The dimension of the unstable subspace satisfies $\dim E^u(L) = 3$.

l) The dimension of the asymptotically stable subspace satisfies $\dim E^s(L) = 3$. □

### 5.3.4 Case 5

The forms of the eigenvalues are $\pm i\beta_j\,(\beta_j \in \mathrm{R}, \beta_j > 0, j=1)$ and $\pm\sigma \pm i\tau\,(\sigma,\tau \in \mathrm{R}; \sigma,\tau > 0)$. The equilibrium point is unstable. The motion of the spacecraft near the equilibrium point relative to the equilibrium point is expressed as

$$\begin{cases} \xi = C_{\xi 1}\cos\beta_1 t + S_{\xi 1}\sin\beta_1 t + E_\xi e^{\sigma_1 t}\cos\tau_1 t + F_\xi e^{\sigma_1 t}\sin\tau_1 t + G_\xi e^{-\sigma_1 t}\cos\tau_1 t + H_\xi e^{-\sigma_1 t}\sin\tau_1 t \\ \eta = C_{\eta 1}\cos\beta_1 t + S_{\eta 1}\sin\beta_1 t + E_\eta e^{\sigma_1 t}\cos\tau_1 t + F_\eta e^{\sigma_1 t}\sin\tau_1 t + G_\eta e^{-\sigma_1 t}\cos\tau_1 t + H_\eta e^{-\sigma_1 t}\sin\tau_1 t \\ \zeta = C_{\zeta 1}\cos\beta_1 t + S_{\zeta 1}\sin\beta_1 t + E_\zeta e^{\sigma_1 t}\cos\tau_1 t + F_\zeta e^{\sigma_1 t}\sin\tau_1 t + G_\zeta e^{-\sigma_1 t}\cos\tau_1 t + H_\zeta e^{-\sigma_1 t}\sin\tau_1 t \end{cases} \quad (35)$$

The almost periodic orbit near the equilibrium point can be expressed as

$$\begin{cases} \xi = C_{\xi 1}\cos\beta_1 t + S_{\xi 1}\sin\beta_1 t + G_\xi e^{-\sigma_1 t}\cos\tau_1 t + H_\xi e^{-\sigma_1 t}\sin\tau_1 t \\ \eta = C_{\eta 1}\cos\beta_1 t + S_{\eta 1}\sin\beta_1 t + G_\eta e^{-\sigma_1 t}\cos\tau_1 t + H_\eta e^{-\sigma_1 t}\sin\tau_1 t \\ \zeta = C_{\zeta 1}\cos\beta_1 t + S_{\zeta 1}\sin\beta_1 t + G_\zeta e^{-\sigma_1 t}\cos\tau_1 t + H_\zeta e^{-\sigma_1 t}\sin\tau_1 t \end{cases} \quad (36)$$

The central manifold is given by

$$\begin{cases} E_\xi = F_\xi = G_\xi = H_\xi = 0 \\ E_\eta = F_\eta = G_\eta = H_\eta = 0 \\ E_\zeta = F_\zeta = G_\zeta = H_\zeta = 0 \end{cases} \quad (37)$$

It is a 2-dimensional smooth manifold.

The asymptotically stable manifold is generated by

$$\begin{cases} \xi = G_\xi e^{-\sigma_1 t}\cos\tau_1 t + H_\xi e^{-\sigma_1 t}\sin\tau_1 t \\ \eta = G_\eta e^{-\sigma_1 t}\cos\tau_1 t + H_\eta e^{-\sigma_1 t}\sin\tau_1 t \\ \zeta = G_\zeta e^{-\sigma_1 t}\cos\tau_1 t + H_\zeta e^{-\sigma_1 t}\sin\tau_1 t \end{cases} \quad (38)$$

It is a 2-dimensional smooth manifold.

The unstable manifold is generated by

$$\begin{cases} \xi = E_\xi e^{\sigma_1 t}\cos\tau_1 t + F_\xi e^{\sigma_1 t}\sin\tau_1 t \\ \eta = E_\eta e^{\sigma_1 t}\cos\tau_1 t + F_\eta e^{\sigma_1 t}\sin\tau_1 t \\ \zeta = E_\zeta e^{\sigma_1 t}\cos\tau_1 t + F_\zeta e^{\sigma_1 t}\sin\tau_1 t \end{cases} \quad (39)$$

It is a 2-dimensional smooth manifold. Then, the general result for Case 5 is stated as follows.

**Theorem 9.** For an equilibrium point in the potential field of a rotating asteroid, the following conditions are equivalent:

a) The roots of the characteristic equation $P(\lambda)$ are in the form of $\pm i\beta_j \left(\beta_j \in \mathrm{R}, \beta_j > 0, j=1\right)$ and $\pm\sigma \pm i\tau \left(\sigma, \tau \in \mathrm{R}; \sigma, \tau > 0\right)$.

b) The motion of the spacecraft near the equilibrium point relative to the equilibrium point can be expressed as

$$\begin{cases} \xi = C_{\xi1} \cos\beta_1 t + S_{\xi1} \sin\beta_1 t + E_\xi e^{\sigma_1 t} \cos\tau_1 t + F_\xi e^{\sigma_1 t} \sin\tau_1 t + G_\xi e^{-\sigma_1 t} \cos\tau_1 t + H_\xi e^{-\sigma_1 t} \sin\tau_1 t \\ \eta = C_{\eta1} \cos\beta_1 t + S_{\eta1} \sin\beta_1 t + E_\eta e^{\sigma_1 t} \cos\tau_1 t + F_\eta e^{\sigma_1 t} \sin\tau_1 t + G_\eta e^{-\sigma_1 t} \cos\tau_1 t + H_\eta e^{-\sigma_1 t} \sin\tau_1 t \\ \zeta = C_{\zeta1} \cos\beta_1 t + S_{\zeta1} \sin\beta_1 t + E_\zeta e^{\sigma_1 t} \cos\tau_1 t + F_\zeta e^{\sigma_1 t} \sin\tau_1 t + G_\zeta e^{-\sigma_1 t} \cos\tau_1 t + H_\zeta e^{-\sigma_1 t} \sin\tau_1 t \end{cases}$$

c) There is only one family of periodic orbits in the tangent space of the equilibrium point, which has the period $T_1 = \dfrac{2\pi}{\beta_1}$, and the dimension of the unstable manifold satisfies $\dim W^u(\mathbf{S}) = 2$. There is no characteristic root in the real axis.

d) There is only one family of periodic orbits near the equilibrium point, and the dimension of the unstable manifold satisfies $\dim W^u(\mathbf{S}) = 2$. There is no characteristic root in the real axis.

e) The asymptotically stable manifold is generated by

$$\begin{cases} \xi = G_\xi e^{-\sigma_1 t} \cos\tau_1 t + H_\xi e^{-\sigma_1 t} \sin\tau_1 t \\ \eta = G_\eta e^{-\sigma_1 t} \cos\tau_1 t + H_\eta e^{-\sigma_1 t} \sin\tau_1 t \\ \zeta = G_\zeta e^{-\sigma_1 t} \cos\tau_1 t + H_\zeta e^{-\sigma_1 t} \sin\tau_1 t \end{cases}.$$

f) The unstable manifold is generated by

$$\begin{cases} \xi = E_\xi e^{\sigma_1 t} \cos\tau_1 t + F_\xi e^{\sigma_1 t} \sin\tau_1 t \\ \eta = E_\eta e^{\sigma_1 t} \cos\tau_1 t + F_\eta e^{\sigma_1 t} \sin\tau_1 t \\ \zeta = E_\zeta e^{\sigma_1 t} \cos\tau_1 t + F_\zeta e^{\sigma_1 t} \sin\tau_1 t \end{cases}.$$

g) The structure of the submanifold is $(\mathbf{S}, \Omega) \simeq T\Xi \cong W^s(\mathbf{S}) \oplus W^c(\mathbf{S}) \oplus W^u(\mathbf{S})$, and $\dim W^s(\mathbf{S}) = \dim W^c(\mathbf{S}) = \dim W^u(\mathbf{S}) = 2$. There is no characteristic root in the real axis.

h) The structure of the subspace is $T_L\mathbf{S} \cong E^s(L) \oplus E^c(L) \oplus E^u(L)$, and $\dim E^s(L) = \dim E^c(L) = \dim E^u(L) = 2$. There is no characteristic root in the real axis. □

**5.4 Resonant Equilibrium Points**

The resonant manifold and the resonant orbits exist if and only if one of Cases 6-8 are established. An equilibrium point is resonant if and only if $\dim W^r(\mathbf{S}) > 0$, with

Case 6 corresponding to $(\mathbf{S}, \Omega) \simeq T\Xi \simeq W^c(\mathbf{S}) \simeq W^r(\mathbf{S})$,

$\dim W^r(\mathbf{S}) = \dim W^c(\mathbf{S}) = 6$;

Case 7 corresponding to $(\mathbf{S}, \Omega) \simeq T\Xi \simeq W^c(\mathbf{S})$, $\dim W^c(\mathbf{S}) = 6$ and

$\dim W^r(\mathbf{S}) = 4$; as well as

Case 8 corresponding to

$(\mathbf{S}, \Omega) \simeq T\Xi \simeq W^s(\mathbf{S}) \oplus W^c(\mathbf{S}) \oplus W^u(\mathbf{S})$, $\dim W^r(\mathbf{S}) = \dim W^c(\mathbf{S}) = 4$, and

$\dim W^s(\mathbf{S}) = \dim W^u(\mathbf{S}) = 1$.

**5.4.1 Case 6**

The motion of the spacecraft relative to the equilibrium point is on a 1:1:1 resonant manifold, which can be expressed as

$$\begin{cases} \xi = C_{\xi 1} \cos \beta_1 t + S_{\xi 1} \sin \beta_1 t + P_{\xi 1} t \cos \beta_1 t + Q_{\xi 1} t \sin \beta_1 t + P_{\xi 2} t^2 \cos \beta_1 t + Q_{\xi 2} t^2 \sin \beta_1 t \\ \eta = C_{\eta 1} \cos \beta_1 t + S_{\eta 1} \sin \beta_1 t + P_{\eta 1} t \cos \beta_1 t + Q_{\eta 1} t \sin \beta_1 t + P_{\eta 2} t^2 \cos \beta_1 t + Q_{\eta 2} t^2 \sin \beta_1 t \\ \zeta = C_{\zeta 1} \cos \beta_1 t + S_{\zeta 1} \sin \beta_1 t + P_{\zeta 1} t \cos \beta_1 t + Q_{\zeta 1} t \sin \beta_1 t + P_{\zeta 2} t^2 \cos \beta_1 t + Q_{\zeta 2} t^2 \sin \beta_1 t \end{cases} \quad (40)$$

The resonant manifold is a 6-dimensional smooth manifold. There is only one family of periodic orbits, which is given by $\begin{cases} P_{\xi 1} = Q_{\xi 1} = P_{\xi 2} = Q_{\xi 2} = 0 \\ P_{\eta 1} = Q_{\eta 1} = P_{\eta 2} = Q_{\eta 2} = 0 \\ P_{\zeta 1} = Q_{\zeta 1} = P_{\zeta 2} = Q_{\zeta 2} = 0 \end{cases}$, and the period is

$T_1 = \dfrac{2\pi}{\beta_1}$. Then, the general result of Case 6 is stated as follows.

**Theorem 10.** For an equilibrium point in the potential field of a rotating asteroid, the following conditions are equivalent:

a) The roots of the characteristic equation $P(\lambda)$ are in the form of

$\pm i\beta_j \left( \beta_j \in \mathrm{R}, \beta_1 = \beta_2 = \beta_3 > 0; j = 1, 2, 3 \right)$.

b) The motion of the spacecraft near the equilibrium point relative to the equilibrium point can be expressed as

$\begin{cases} \xi = C_{\xi 1} \cos \beta_1 t + S_{\xi 1} \sin \beta_1 t + P_{\xi 1} t \cos \beta_1 t + Q_{\xi 1} t \sin \beta_1 t + P_{\xi 2} t^2 \cos \beta_1 t + Q_{\xi 2} t^2 \sin \beta_1 t \\ \eta = C_{\eta 1} \cos \beta_1 t + S_{\eta 1} \sin \beta_1 t + P_{\eta 1} t \cos \beta_1 t + Q_{\eta 1} t \sin \beta_1 t + P_{\eta 2} t^2 \cos \beta_1 t + Q_{\eta 2} t^2 \sin \beta_1 t \\ \zeta = C_{\zeta 1} \cos \beta_1 t + S_{\zeta 1} \sin \beta_1 t + P_{\zeta 1} t \cos \beta_1 t + Q_{\zeta 1} t \sin \beta_1 t + P_{\zeta 2} t^2 \cos \beta_1 t + Q_{\zeta 2} t^2 \sin \beta_1 t \end{cases}$

c) The structure of the submanifold is $(\mathbf{S}, \Omega) \simeq T\Xi \cong W^c(\mathbf{S}) \cong W^r(\mathbf{S})$.

d) The resonant manifold is a 6-dimensional manifold: $\dim W^r(\mathbf{S}) = \dim W^c(\mathbf{S}) = 6$.

e) The structure of the subspace is $T_L \mathbf{S} \cong E^c(L) \cong E^r(L)$.

f) The resonant subspace is a 6-dimensional space: $\dim E^c(L) = \dim E^r(L) = 6$. □

### 5.4.2 Case 7

The motion of the spacecraft near the equilibrium point relative to the equilibrium point is expressed as

$$\begin{cases} \xi = C_{\xi 1}\cos\beta_1 t + S_{\xi 1}\sin\beta_1 t + P_{\xi 1}t\cos\beta_1 t + Q_{\xi 1}t\sin\beta_1 t + C_{\xi 3}\cos\beta_3 t + S_{\xi 3}\sin\beta_3 t \\ \eta = C_{\eta 1}\cos\beta_1 t + S_{\eta 1}\sin\beta_1 t + P_{\eta 1}t\cos\beta_1 t + Q_{\eta 1}t\sin\beta_1 t + C_{\eta 3}\cos\beta_3 t + S_{\eta 3}\sin\beta_3 t \\ \zeta = C_{\zeta 1}\cos\beta_1 t + S_{\zeta 1}\sin\beta_1 t + P_{\zeta 1}t\cos\beta_1 t + Q_{\zeta 1}t\sin\beta_1 t + C_{\zeta 3}\cos\beta_3 t + S_{\zeta 3}\sin\beta_3 t \end{cases} \quad (41)$$

There is a 1:1 resonant manifold, which is generated by

$$\begin{cases} \xi = C_{\xi 1}\cos\beta_1 t + S_{\xi 1}\sin\beta_1 t + P_{\xi 1}t\cos\beta_1 t + Q_{\xi 1}t\sin\beta_1 t \\ \eta = C_{\eta 1}\cos\beta_1 t + S_{\eta 1}\sin\beta_1 t + P_{\eta 1}t\cos\beta_1 t + Q_{\eta 1}t\sin\beta_1 t \\ \zeta = C_{\zeta 1}\cos\beta_1 t + S_{\zeta 1}\sin\beta_1 t + P_{\zeta 1}t\cos\beta_1 t + Q_{\zeta 1}t\sin\beta_1 t \end{cases} \quad (42)$$

The resonant manifold is a 4-dimensional smooth manifold. There are two families of periodic orbits. The first family of periodic orbits is given by

$$\begin{cases} P_{\xi 1} = Q_{\xi 1} = C_{\xi 3} = S_{\xi 3} = 0 \\ P_{\eta 1} = Q_{\eta 1} = C_{\eta 3} = S_{\eta 3} = 0 \\ P_{\zeta 1} = Q_{\zeta 1} = C_{\zeta 3} = S_{\zeta 3} = 0 \end{cases}$$, and the period is $T_1 = \dfrac{2\pi}{\beta_1}$. The second family of periodic

orbits is given by $\begin{cases} P_{\xi 1} = Q_{\xi 1} = C_{\xi 1} = S_{\xi 1} = 0 \\ P_{\eta 1} = Q_{\eta 1} = C_{\eta 1} = S_{\eta 1} = 0 \\ P_{\zeta 1} = Q_{\zeta 1} = C_{\zeta 1} = S_{\zeta 1} = 0 \end{cases}$, and the period is $T_3 = \dfrac{2\pi}{\beta_3}$. Then, the general result of Case 7 is stated as follows.

**Theorem 11.** For an equilibrium point in the potential field of a rotating asteroid, the following conditions are equivalent:

a) The roots of the characteristic equation $P(\lambda)$ are in the form of $\pm i\beta_j \left(\beta_j \in \mathrm{R}, \beta_j > 0, \beta_1 = \beta_2 \neq \beta_3; j = 1,2,3\right)$.

b) The motion of the spacecraft near the equilibrium point relative to the equilibrium point can be expressed as

$$\begin{cases} \xi = C_{\xi 1}\cos\beta_1 t + S_{\xi 1}\sin\beta_1 t + P_{\xi 1}t\cos\beta_1 t + Q_{\xi 1}t\sin\beta_1 t + C_{\xi 3}\cos\beta_3 t + S_{\xi 3}\sin\beta_3 t \\ \eta = C_{\eta 1}\cos\beta_1 t + S_{\eta 1}\sin\beta_1 t + P_{\eta 1}t\cos\beta_1 t + Q_{\eta 1}t\sin\beta_1 t + C_{\eta 3}\cos\beta_3 t + S_{\eta 3}\sin\beta_3 t \\ \zeta = C_{\zeta 1}\cos\beta_1 t + S_{\zeta 1}\sin\beta_1 t + P_{\zeta 1}t\cos\beta_1 t + Q_{\zeta 1}t\sin\beta_1 t + C_{\zeta 3}\cos\beta_3 t + S_{\zeta 3}\sin\beta_3 t \end{cases}$$

c) There are two families of periodic orbits in the tangent space of the equilibrium point, and $\dim W^r(\mathbf{S}) = 4$.

d) The structure of the submanifold is $(S,\Omega) \simeq T\Xi \cong W^c(S)$, and $\dim W^r(S) = 4$.

e) $\dim W^c(S) = 6$ and $\dim W^r(S) = 4$.

f) The structure of the subspace is $T_L S \cong E^c(L)$, and $\dim E^r(L) = 4$.

g) $\dim E^c(L) = 6$ and $\dim E^r(L) = 4$. □

### 5.4.3 Case 8

The motion of the spacecraft near the equilibrium point relative to the equilibrium point is expressed as

$$\begin{cases} \xi = A_{\xi 1}e^{\alpha_1 t} + B_{\xi 1}e^{-\alpha_1 t} + C_{\xi 1}\cos\beta_1 t + S_{\xi 1}\sin\beta_1 t + P_{\xi 1}t\cos\beta_1 t + Q_{\xi 1}t\sin\beta_1 t \\ \eta = A_{\eta 1}e^{\alpha_1 t} + B_{\eta 1}e^{-\alpha_1 t} + C_{\eta 1}\cos\beta_1 t + S_{\eta 1}\sin\beta_1 t + P_{\eta 1}t\cos\beta_1 t + Q_{\eta 1}t\sin\beta_1 t \\ \zeta = A_{\zeta 1}e^{\alpha_1 t} + B_{\zeta 1}e^{-\alpha_1 t} + C_{\zeta 1}\cos\beta_1 t + S_{\zeta 1}\sin\beta_1 t + P_{\zeta 1}t\cos\beta_1 t + Q_{\zeta 1}t\sin\beta_1 t \end{cases} \quad (43)$$

There is a 1:1 resonant manifold, which is generated by

$$\begin{cases} \xi = C_{\xi 1}\cos\beta_1 t + S_{\xi 1}\sin\beta_1 t + P_{\xi 1}t\cos\beta_1 t + Q_{\xi 1}t\sin\beta_1 t \\ \eta = C_{\eta 1}\cos\beta_1 t + S_{\eta 1}\sin\beta_1 t + P_{\eta 1}t\cos\beta_1 t + Q_{\eta 1}t\sin\beta_1 t \\ \zeta = C_{\zeta 1}\cos\beta_1 t + S_{\zeta 1}\sin\beta_1 t + P_{\zeta 1}t\cos\beta_1 t + Q_{\zeta 1}t\sin\beta_1 t \end{cases} \quad (44)$$

The resonant manifold is a 4-dimensional smooth manifold. There is only one family of periodic orbits, which is given by $\begin{cases} A_{\xi 1} = B_{\xi 1} = P_{\xi 1} = Q_{\xi 1} = 0 \\ A_{\eta 1} = B_{\eta 1} = P_{\eta 1} = Q_{\eta 1} = 0 \\ A_{\zeta 1} = B_{\zeta 1} = P_{\zeta 1} = Q_{\zeta 1} = 0 \end{cases}$, and the period is

$T_1 = \dfrac{2\pi}{\beta_1}$. The asymptotically stable manifold is generated by

$$\begin{cases} \xi = B_{\xi 1}e^{-\alpha_1 t} \\ \eta = B_{\eta 1}e^{-\alpha_1 t} \\ \zeta = B_{\zeta 1}e^{-\alpha_1 t} \end{cases} \quad (45)$$

It is a 1-dimensional smooth manifold. The unstable manifold is generated by

$$\begin{cases} \xi = A_{\xi 1} e^{\alpha_1 t} \\ \eta = A_{\eta 1} e^{\alpha_1 t} \\ \zeta = A_{\zeta 1} e^{\alpha_1 t} \end{cases} \tag{46}$$

It is a 1-dimensional smooth manifold. Then, the general result of Case 8 is stated as follows.

**Theorem 12.** For an equilibrium point in the potential field of a rotating asteroid, the following conditions are equivalent:

a) The roots of the characteristic equation $P(\lambda)$ are in the form of $\pm \alpha_j \left( \alpha_j \in \mathrm{R}, \alpha_j > 0, j = 1 \right)$ and $\pm i\beta_j \left( \beta_j \in \mathrm{R}, \beta_1 = \beta_2 > 0; j = 1, 2 \right)$.

b) The motion of the spacecraft near the equilibrium point relative to the equilibrium point can be expressed as

$$\begin{cases} \xi = A_{\xi 1} e^{\alpha_1 t} + B_{\xi 1} e^{-\alpha_1 t} + C_{\xi 1} \cos \beta_1 t + S_{\xi 1} \sin \beta_1 t + P_{\xi 1} t \cos \beta_1 t + Q_{\xi 1} t \sin \beta_1 t \\ \eta = A_{\eta 1} e^{\alpha_1 t} + B_{\eta 1} e^{-\alpha_1 t} + C_{\eta 1} \cos \beta_1 t + S_{\eta 1} \sin \beta_1 t + P_{\eta 1} t \cos \beta_1 t + Q_{\eta 1} t \sin \beta_1 t \\ \zeta = A_{\zeta 1} e^{\alpha_1 t} + B_{\zeta 1} e^{-\alpha_1 t} + C_{\zeta 1} \cos \beta_1 t + S_{\zeta 1} \sin \beta_1 t + P_{\zeta 1} t \cos \beta_1 t + Q_{\zeta 1} t \sin \beta_1 t \end{cases}$$

c) There is one family of periodic orbits in the tangent space of the equilibrium point, and $\dim W^r(\mathbf{S}) = 4$.

d) The structure of the submanifold is $(\mathbf{S}, \Omega) \simeq T\Xi \cong W^s(\mathbf{S}) \oplus W^c(\mathbf{S}) \oplus W^u(\mathbf{S})$, and $\dim W^r(\mathbf{S}) = 4$.

e) $\dim W^c(\mathbf{S}) = \dim W^r(\mathbf{S}) = 4$.

f) $\dim W^c(\mathbf{S}) = \dim W^r(\mathbf{S}) = 4$ and $\dim W^s(\mathbf{S}) = \dim W^u(\mathbf{S}) = 1$.

g) The structure of the subspace is $T_L \mathbf{S} \cong E^s(L) \oplus E^c(L) \oplus E^u(L)$, and $\dim E^r(L) = 4$.

h) $\dim E^c(L) = \dim E^r(L) = 4$ and $\dim E^s(L) = \dim E^u(L) = 1$. □

## 6. Applications to Asteroids

In this section, the theorems described in the previous sections are applied to asteroids 216 Kleopatra, 1620 Geographos, 4769 Castalia, and 6489 Golevka. The physical model of these four asteroids that we used here was calculated with radar observations using the polyhedral model (Neese 2004).

### 6.1 Application to Asteroid 216 Kleopatra

The estimated bulk density of asteroid 216 Kleopatra is 3.6 $g \cdot cm^{-3}$ (Descamps et al. 2011), and the rotation period of asteroid 216 Kleopatra is 5.385 h (Ostro et al. 2000). Table 1 shows the positions of the equilibrium points in the body-fixed frame, which were calculated by Newton method using Eq. (9). Table 2 shows the eigenvalues of the equilibrium points, which were calculated using Eq. (14). E1 and E2 belong to Case 2, whereas E3 and E4 belong to Case 5. Yu & Baoyin (2012a) calculated the positions of the equilibrium points as well as the eigenvalues of the equilibrium points for asteroid 216 Kleopatra using the numerical method.

Table 1 Positions of the Equilibrium Points around Asteroid 216 Kleopatra

| Equilibrium Points | x (km) | y (km) | z (km) |
|---|---|---|---|
| E1 | 142. 852 | 2.45436 | 1.18008 |
| E2 | -144.684 | 5.18855 | -0.282998 |
| E3 | 2.21701 | -102.102 | 0.279703 |
| E4 | -1.16396 | 100.738 | -0.541516 |

Table 2 Eigenvalues of the Equilibrium Points around Asteroid 216 Kleopatra

| $\times 10^{-3}$ | $\lambda_1$ | $\lambda_2$ | $\lambda_3$ | $\lambda_4$ | $\lambda_5$ | $\lambda_6$ |
|---|---|---|---|---|---|---|
| E1 | 0.376 | -0.376 | 0.413i | -0.413i | 0.425i | -0.425i |
| E2 | 0.422 | -0.422 | 0.414i | -0.414i | 0.466i | -0.466i |
| E3 | 0.327i | -0.327i | 0.202+0.304i | 0.202-0.304i | -0.202+0.304i | -0.202-0.304i |
| E4 | 0.323i | -0.323i | 0.202+0.306i | 0.202-0.306i | -0.202+0.306i | -0.202-0.306i |

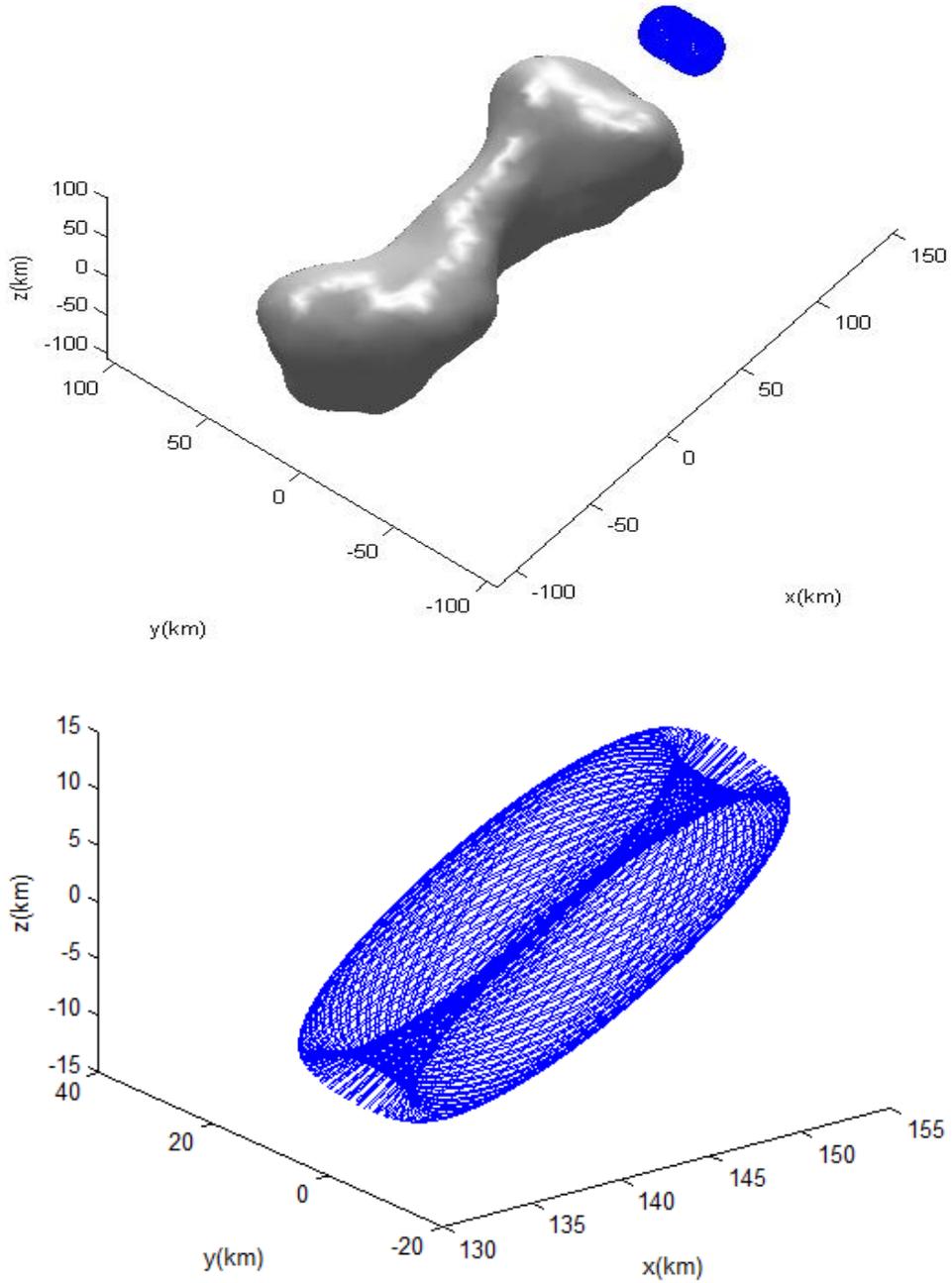

Figure 1.a    A Quasi-Periodic Orbit near the Equilibrium Point E1

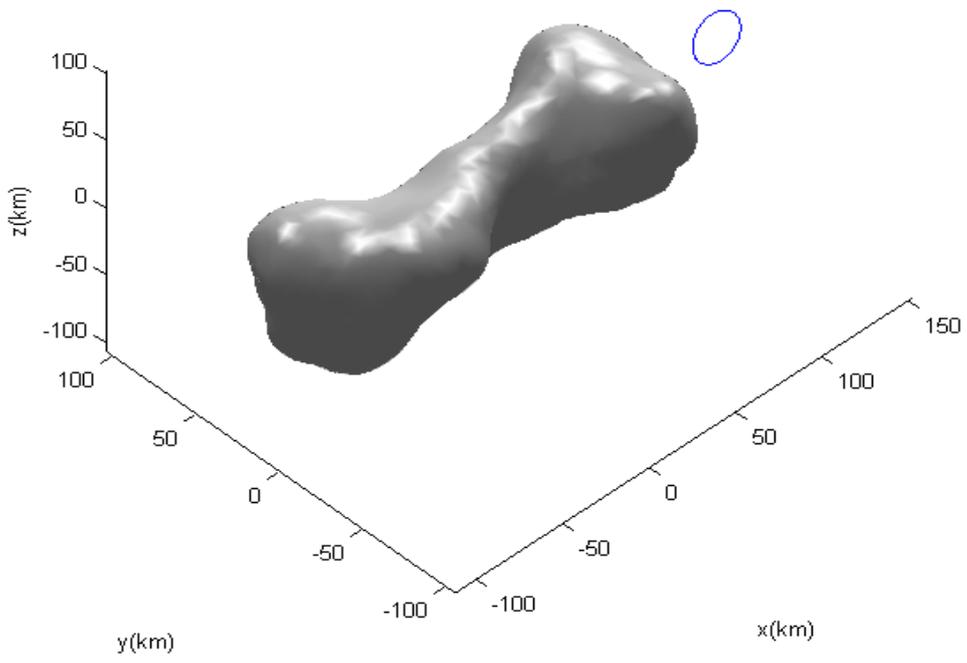

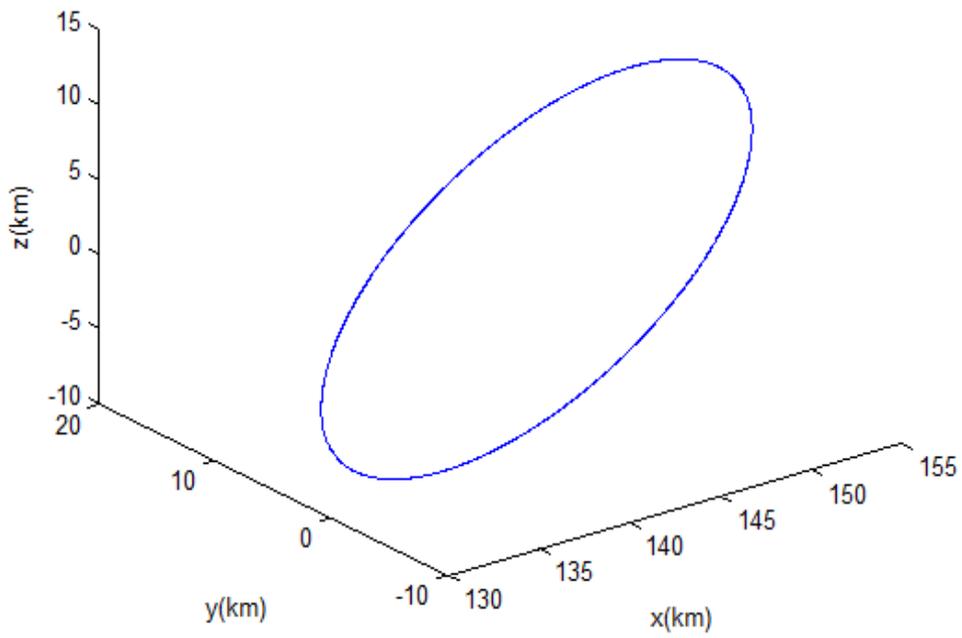

Figure 1.b  A Periodic Orbit near the Equilibrium Point E1

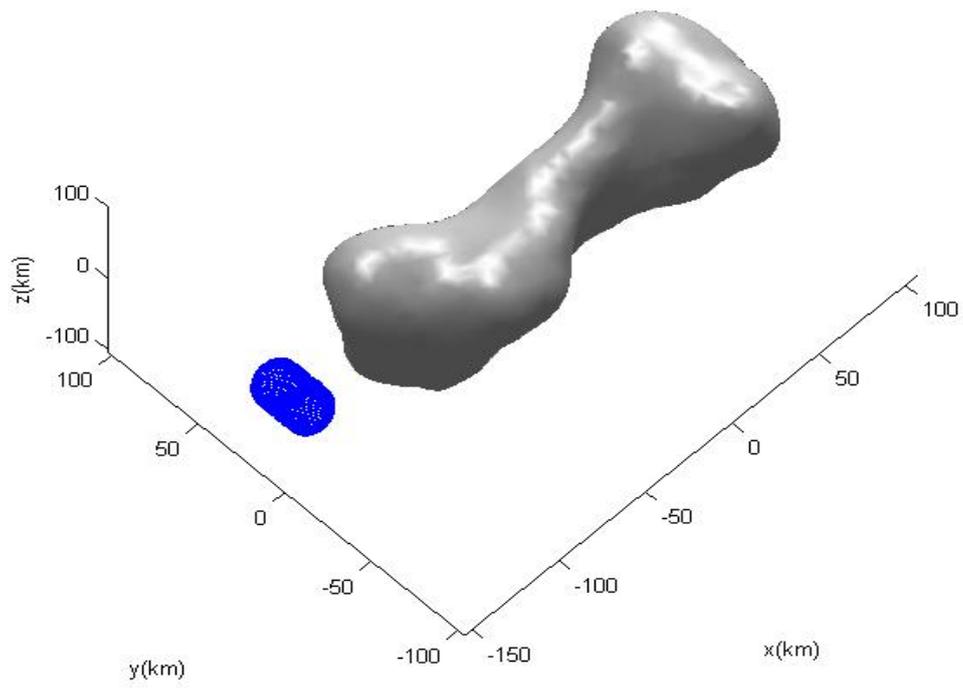

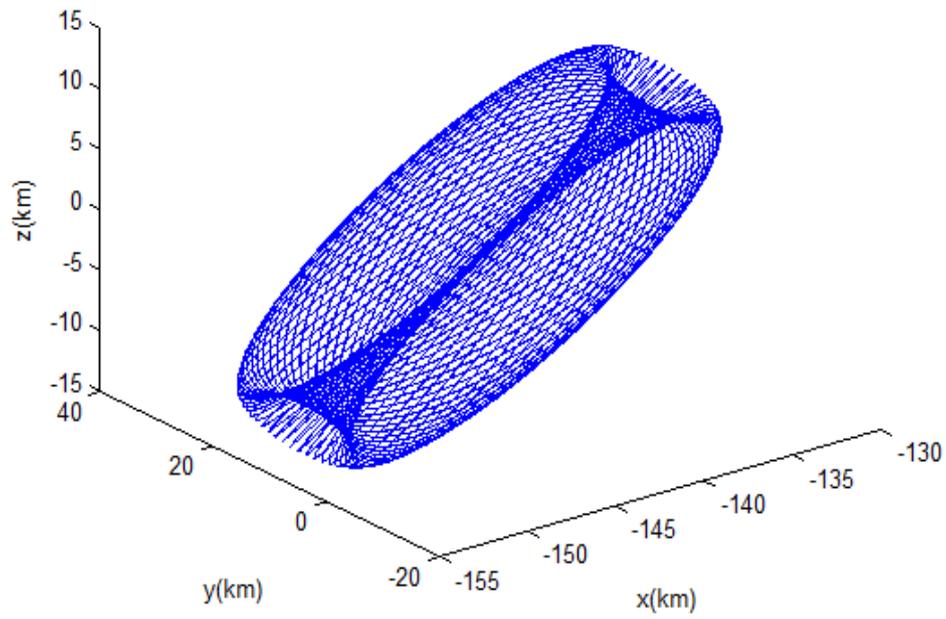

Figure 2.    A Quasi-Periodic Orbit near the Equilibrium Point E2

Figure 1.a shows a quasi-periodic orbit near the equilibrium point E1, where the coefficients have the values $\begin{cases} C_{\xi 1} = C_{\zeta 1} = S_{\eta 1} = C_{\eta 2} = 10 \\ S_{\zeta 2} = 2 \end{cases}$ and other coefficients being equal to zero. The flight time of the orbit is 12 days. Figure 1.b shows a periodic orbit near the equilibrium point E1, where the coefficients have the values $C_{\xi 1} = C_{\zeta 1} = S_{\eta 1} = 10$ and other coefficients being equal to zero. E1 belongs to Case 2. There is one family of quasi-periodic orbits near E1, which is on the 2-dimensional tori $T^2$. The orbit in figure 1 belongs to the quasi-periodic orbital family. For the equilibrium point E2, figure 2 shows a quasi-periodic orbit near the equilibrium point E2, the coefficients have the values $\begin{cases} C_{\xi 1} = C_{\zeta 1} = S_{\eta 1} = C_{\eta 2} = 10 \\ S_{\zeta 2} = 2 \end{cases}$ and other coefficients being equal to zero. The flight time of the orbit is 12 days. E2 also belongs to Case 2. There is one family of quasi-periodic orbits near E2, which is on the 2- dimensional tori $T^2$. The orbit belongs a quasi-periodic orbital family that is notably similar to that of E1.

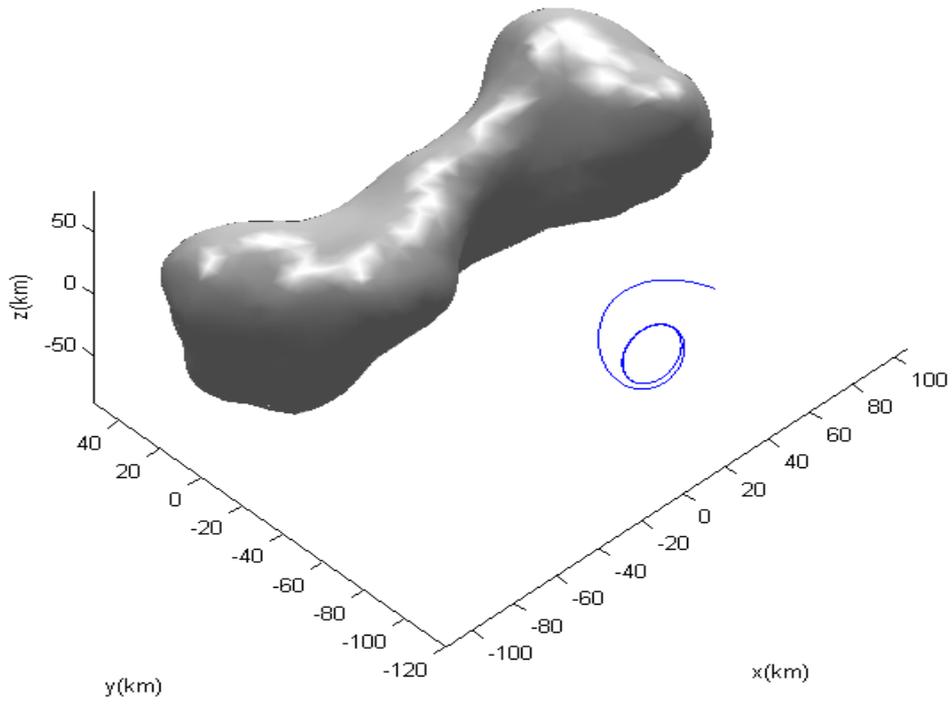

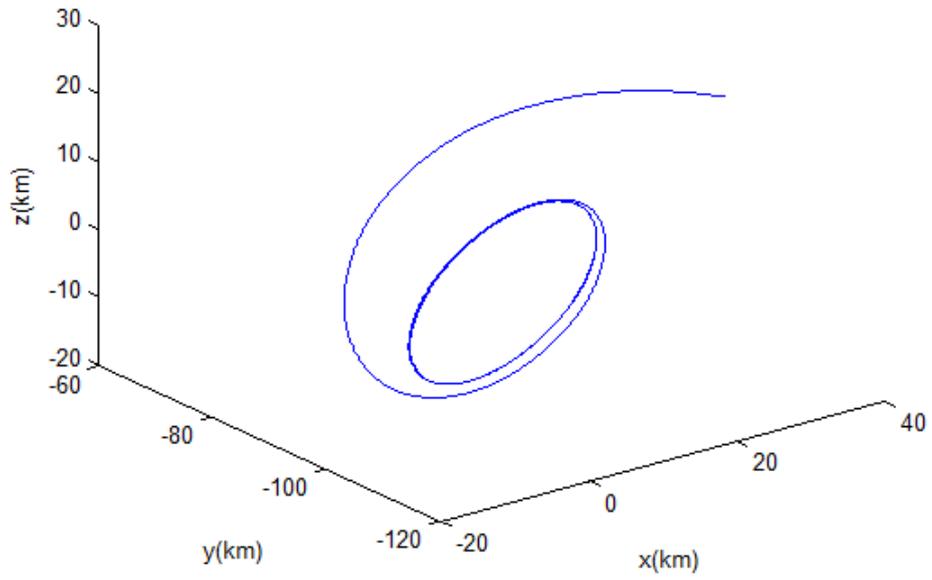

Figure 3.   An Orbit near the Equilibrium Point E3

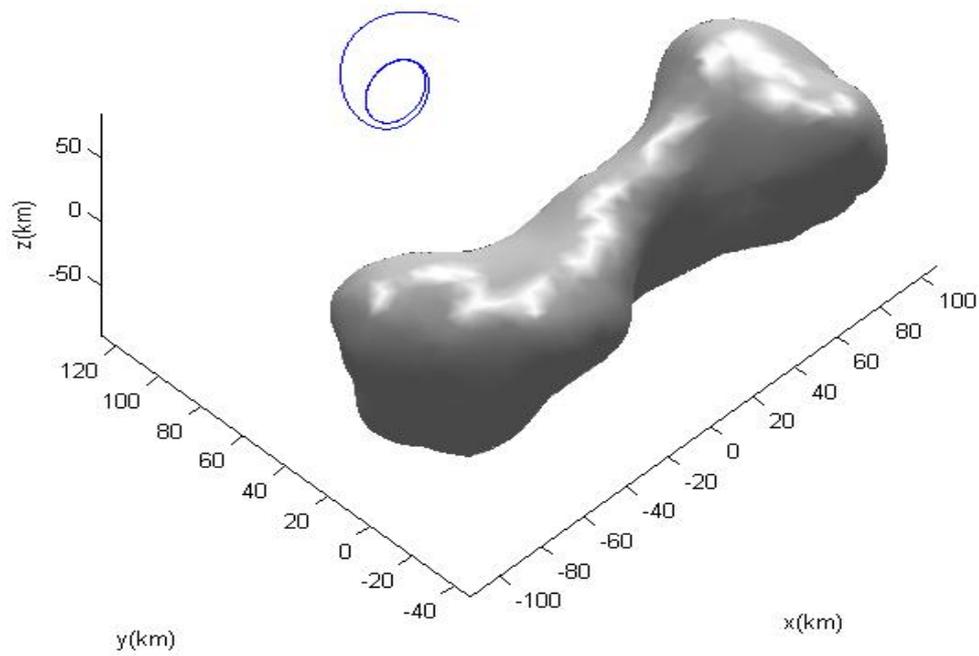

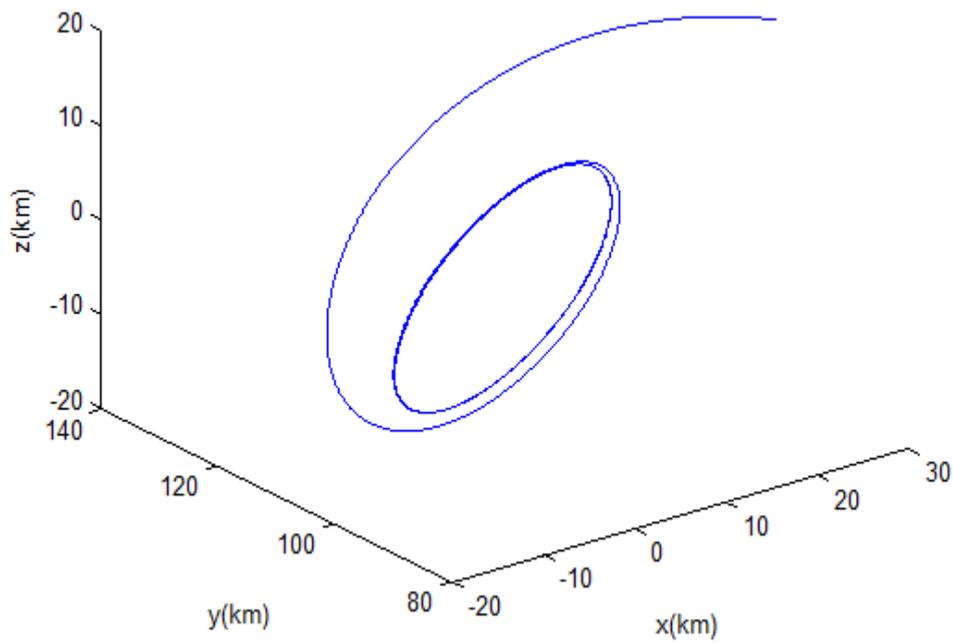

Figure 4.　An Orbit near the Equilibrium Point E4

Figure 3 shows an almost periodic orbit near the equilibrium point E3. The flight time of the orbit is 12 days. There is only one family of periodic orbits near E3.

Figure 4 shows an almost periodic orbit near the equilibrium point E4. The flight time of the orbit around the equilibrium point E4 is 12 days. There is only one family of periodic orbits near E4.

**6.2 Application to Asteroid 1620 Geographos**

The estimated bulk density of asteroid 1620 Geographos is 2.0 $g \cdot cm^{-3}$ (Hudson & Ostro 1999), and the rotation period of asteroid 1620 Geographos is 5.222h (Ryabova 2002). Table 3 shows the positions of the equilibrium points, which were calculated using Eq. (9). Table 4 shows the eigenvalues of the equilibrium points. E1 and E2 belong to Case 2, whereas E3 and E4 belong to Case 5.

Table 3 Positions of the Equilibrium Points around Asteroid 1620 Geographos

| Equilibrium Points | x (km) | y (km) | z (km) |
|---|---|---|---|
| E1 | 2.69925 | -0.041494 | 0.085296 |
| E2 | -2.84097 | -0.057621 | 0.142056 |
| E3 | -0.141618 | 2.11961 | -0.021510 |
| E4 | -0.125678 | -2.08723 | -0.025536 |

Table 4 Eigenvalues of the Equilibrium Points around Asteroid 1620 Geographos

| $\times 10^{-3}$ | $\lambda_1$ | $\lambda_2$ | $\lambda_3$ | $\lambda_4$ | $\lambda_5$ | $\lambda_6$ |
|---|---|---|---|---|---|---|
| E1 | 0.455 | -0.455 | 0.427i | -0.427i | 0.487i | -0.487i |
| E2 | 0.608 | -0.608 | 0.511i | -0.511i | 0.566i | -0.566i |
| E3 | 0.334i | -0.334i | 0.152+0.271i | 0.152-0.271i | -0.152+0.271i | -0.152-0.271i |
| E4 | 0.334i | -0.334i | 0.174+0.284i | 0.174-0.284i | -0.174+0.284i | -0.174-0.284i |

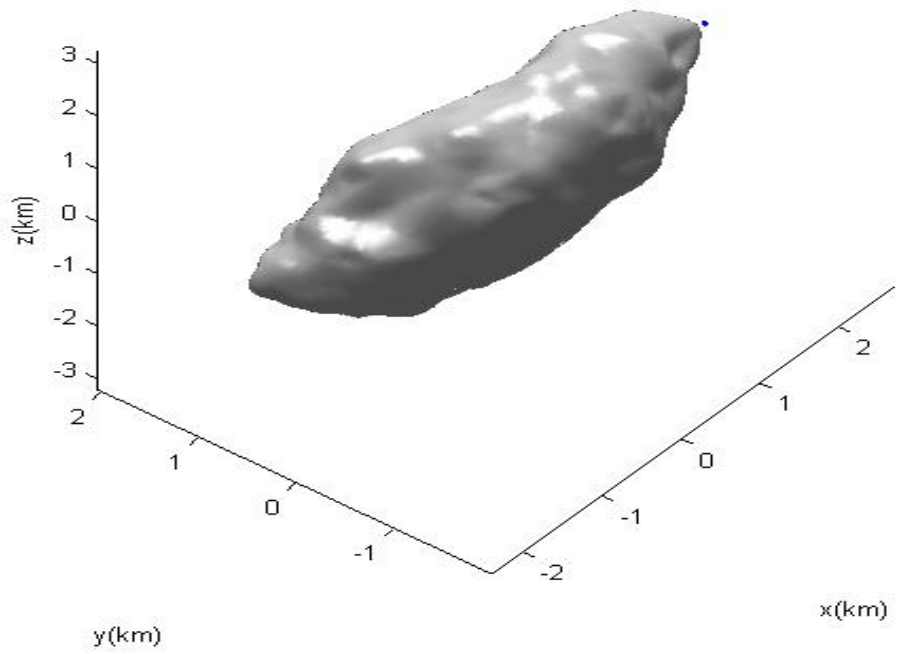

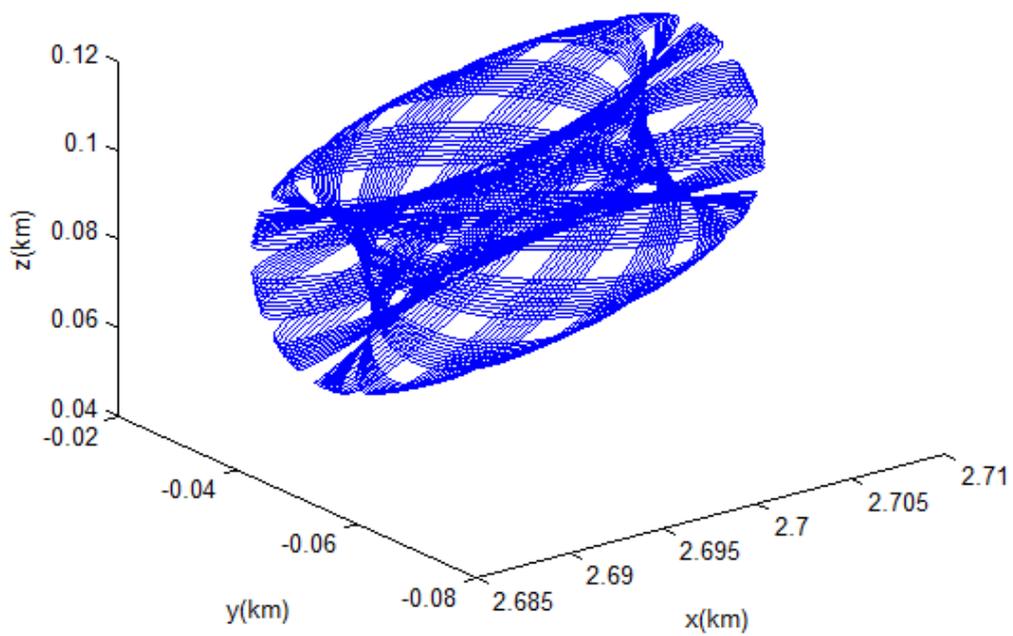

Figure 5.　　A Quasi-Periodic Orbit near the Equilibrium Point E1

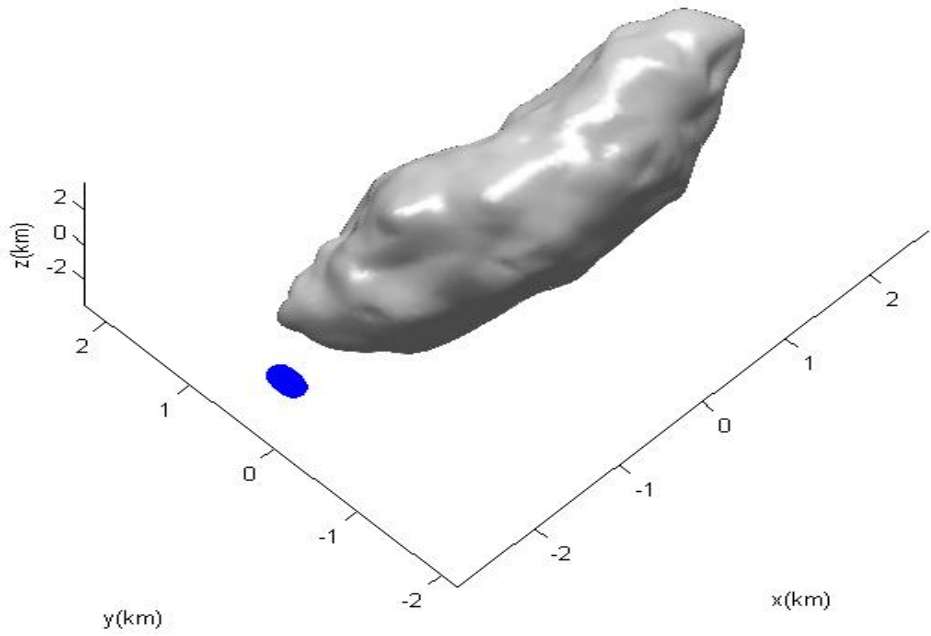

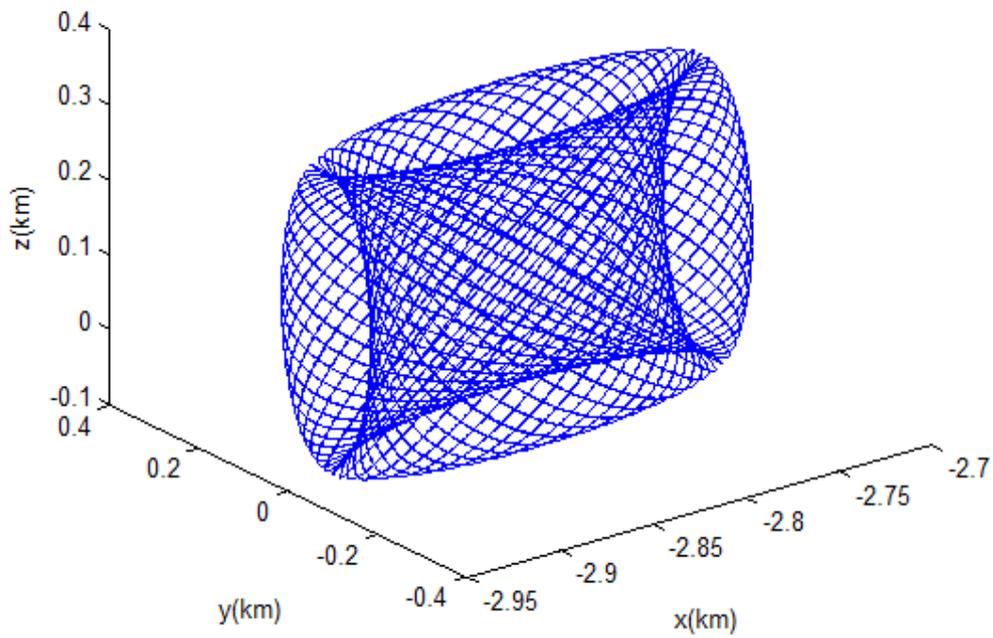

Figure 6. A Quasi-Periodic Orbit near the Equilibrium Point E2

Figure 5 shows a quasi-periodic orbit near the equilibrium point E1, where the coefficients have the values $\begin{cases} C_{\xi 1} = C_{\zeta 1} = S_{\eta 1} = C_{\eta 2} = 0.01 \\ S_{\zeta 2} = 0.02 \end{cases}$ and other coefficients being equal to zero. The flight time of the orbit is 12 days. E1 belongs to Case 2. There is one family of quasi-periodic orbits near E1, which is on the 2-dimensional tori $T^2$. The orbit in figure 5 belongs to the quasi-periodic orbital family. For the equilibrium point E2, figure 6 shows a quasi-periodic orbit near the equilibrium point E2, the coefficients have the values $\begin{cases} C_{\xi 1} = C_{\zeta 1} = S_{\eta 1} = C_{\eta 2} = 0.1 \\ S_{\zeta 2} = 0.2 \end{cases}$ and other coefficients being equal to zero. The flight time of the orbit is 12 days. E2 also belongs to Case 2. There is one family of quasi-periodic orbits near E2, which is on the 2- dimensional tori $T^2$. The orbit belongs a quasi-periodic orbital family that is notably similar to that of E1.

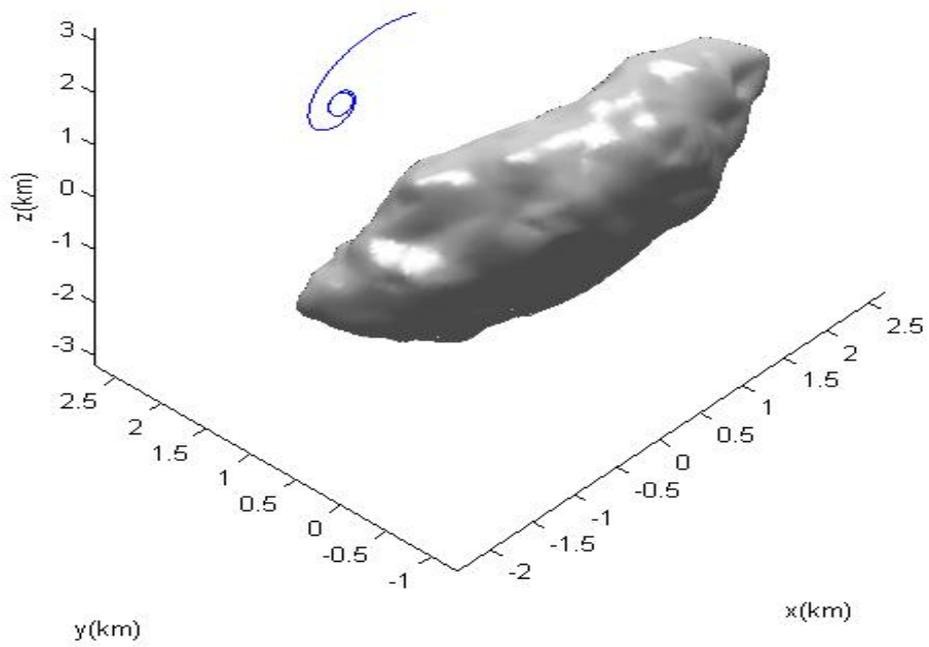

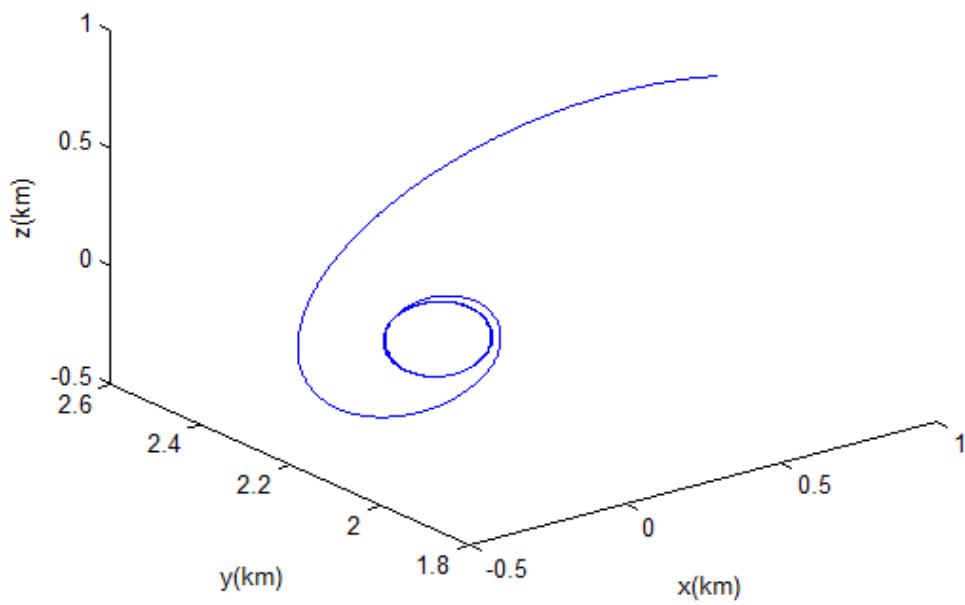

Figure 7.　An Orbit near the Equilibrium Point E3

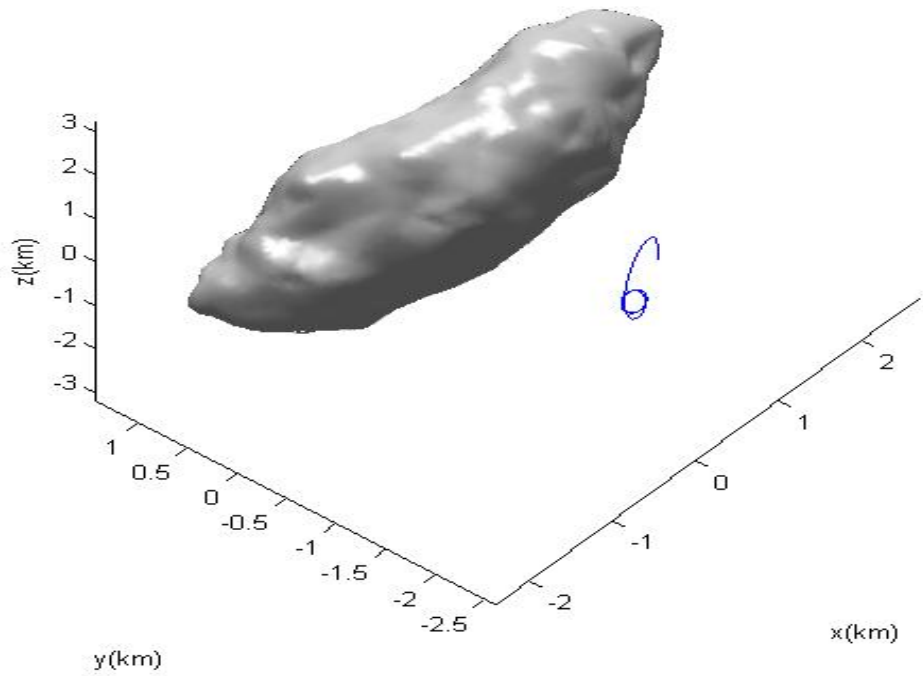

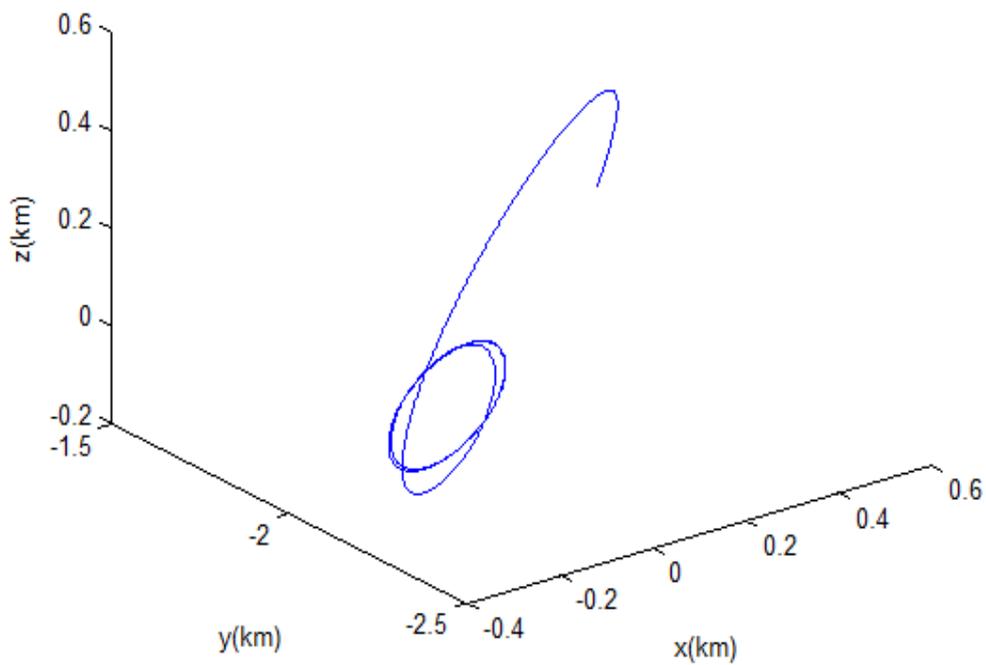

Figure 8. An Orbit near the Equilibrium Point E4

Figure 7 shows an almost periodic orbit near the equilibrium point E3. The flight time of the orbit is 12 days. There is only one family of periodic orbits near E3. Figure 8 shows an almost periodic orbit near the equilibrium point E4. The flight time of the orbit around the equilibrium point E4 is 12 days. There is only one family of periodic orbits near E4.

**6.3 Application to Asteroid 4769 Castalia**

The estimated bulk density of asteroid 4769 Castalia is 2.1 $g \cdot cm^{-3}$ (Hudson & Ostro 1994; Scheeres et al. 1996), and the rotation period of asteroid 4769 Castalia is 4.095h (Hudson et al. 1997). Table 5 shows the positions of the equilibrium points, which were calculated using Eq. (9). Table 6 shows the eigenvalues of the equilibrium points. E1 and E2 belong to Case 2, whereas E3 and E4 belong to Case 5.

Table 5 Positions of the Equilibrium Points around Asteroid 4769 Castalia

| Equilibrium Points | x (km) | y (km) | z (km) |
|---|---|---|---|
| E1 | 0.978767 | 0.023603 | 0.028032 |
| E2 | -1.01550 | 0.116078 | 0.0257862 |
| E3 | -0.043617 | 0.815747 | 0.002265 |
| E4 | -0.034338 | -0.823895 | -0.006950 |

Table 6 Eigenvalues of the Equilibrium Points around Asteroid 4769 Castalia

| $\times 10^{-3}$ | $\lambda_1$ | $\lambda_2$ | $\lambda_3$ | $\lambda_4$ | $\lambda_5$ | $\lambda_6$ |
|---|---|---|---|---|---|---|
| E1 | 0.341 | -0.341 | 0.416i | -0.416i | 0.470i | -0.470i |
| E2 | 0.423 | -0.423 | 0.467i | -0.467i | 0.490i | -0.490i |
| E3 | 0.380i | -0.380i | 0.195+0.324i | 0.195-0.324i | -0.195+0.324i | -0.195-0.324i |
| E4 | 0.386i | -0.386i | 0.197+0.322i | 0.197-0.322i | -0.197+0.322i | -0.197-0.322i |

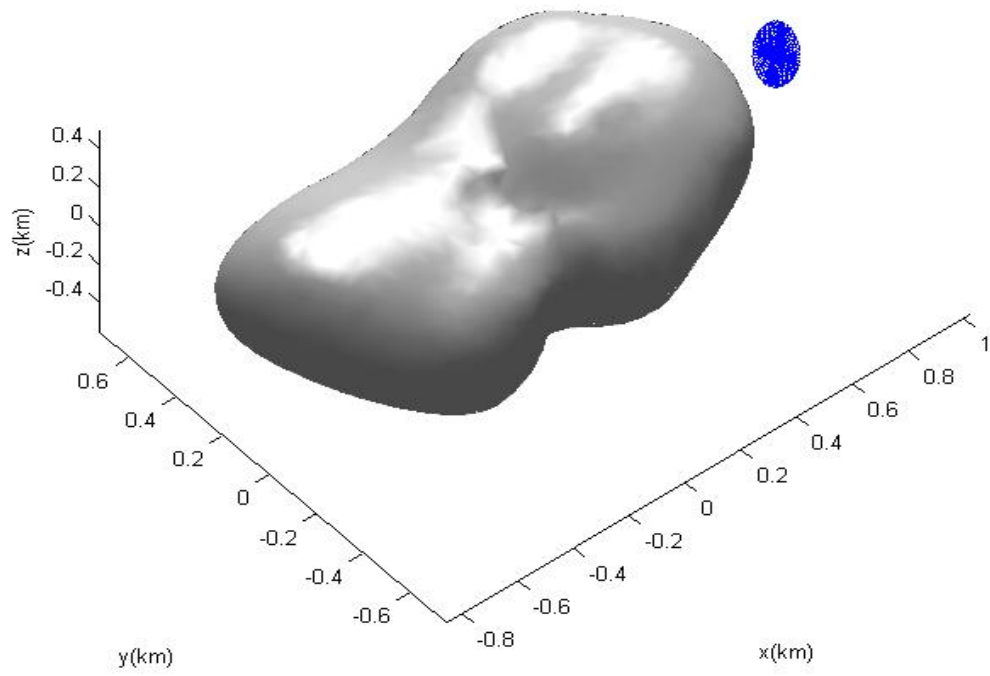

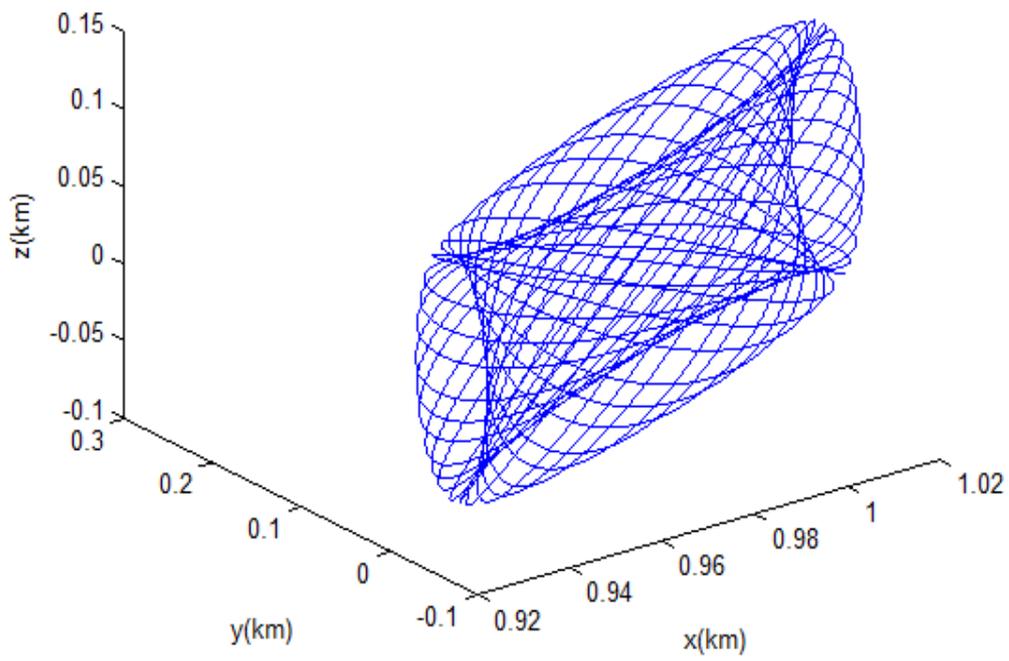

Figure 9.　A Quasi-Periodic Orbit near the Equilibrium Point E1

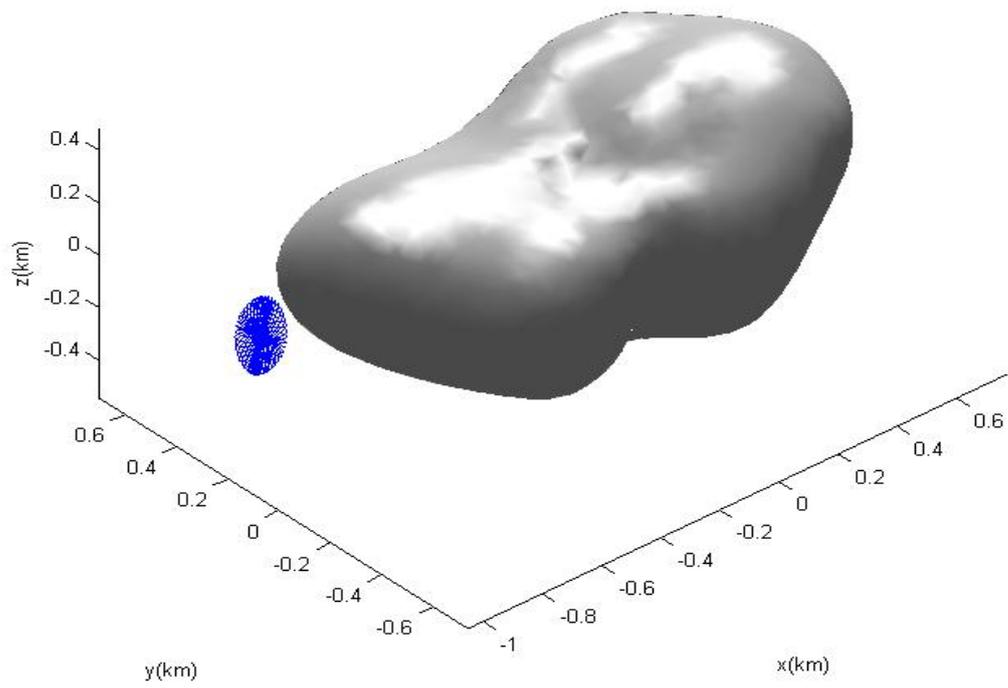

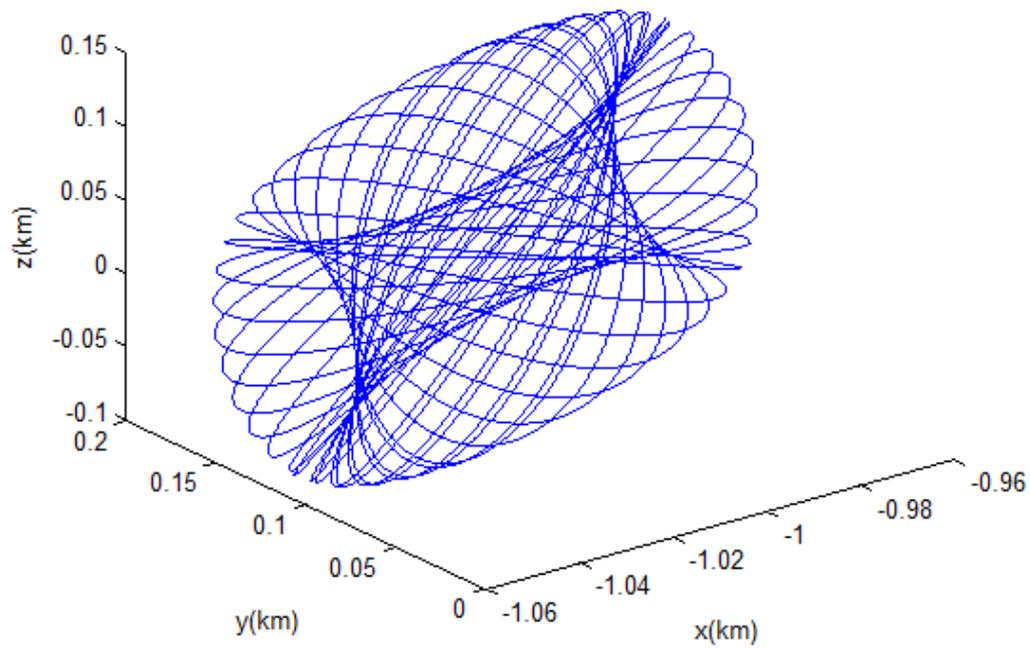

Figure 10. A Quasi-Periodic Orbit near the Equilibrium Point E2

Figure 9 shows a quasi-periodic orbit near the equilibrium point E1, where the coefficients have the values $\begin{cases} C_{\xi 1} = C_{\zeta 1} = S_{\eta 1} = C_{\eta 2} = 0.04 \\ S_{\zeta 2} = 0.08 \end{cases}$ and other coefficients being equal to zero. The flight time of the orbit is 4 days. E1 belongs to Case 2. There is one family of quasi-periodic orbits near E1, which is on the 2-dimensional tori $T^2$. The orbit in figure 9 belongs to the quasi-periodic orbital family. For the equilibrium point E2, figure 10 shows a quasi-periodic orbit near the equilibrium point E2, the coefficients have the values $\begin{cases} C_{\xi 1} = C_{\zeta 1} = S_{\eta 1} = C_{\eta 2} = 0.04 \\ S_{\zeta 2} = 0.08 \end{cases}$ and other coefficients being equal to zero. The flight time of the orbit is 4 days. E2 also belongs to Case 2. There is one family of quasi-periodic orbits near E2, which is on the 2- dimensional tori $T^2$. The orbit belongs a quasi-periodic orbital family that is notably similar to that of E1.

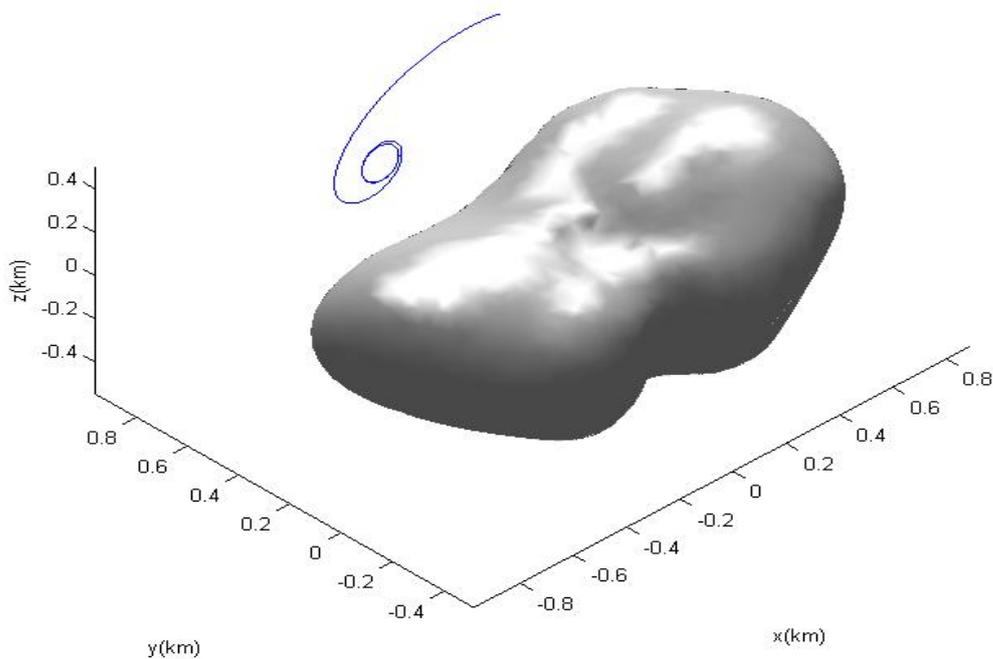

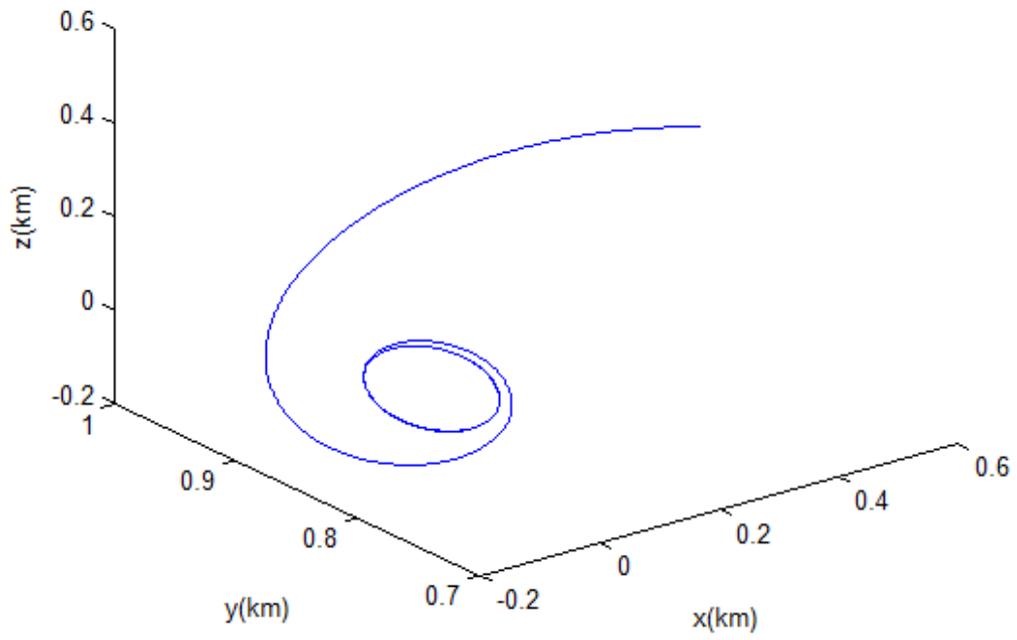

Figure 11. An Orbit near the Equilibrium Point E3

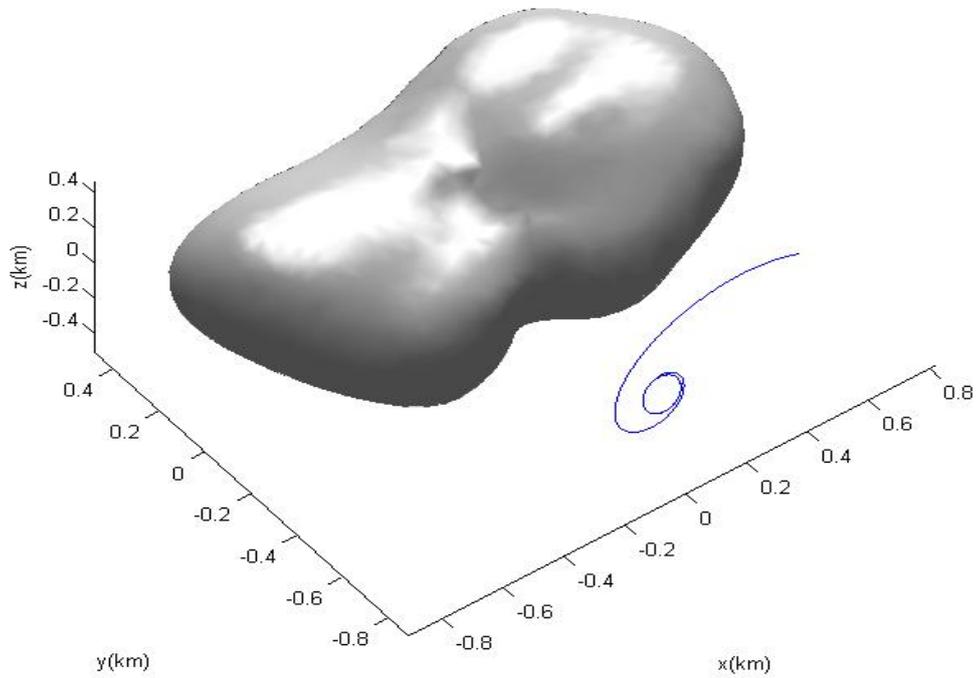

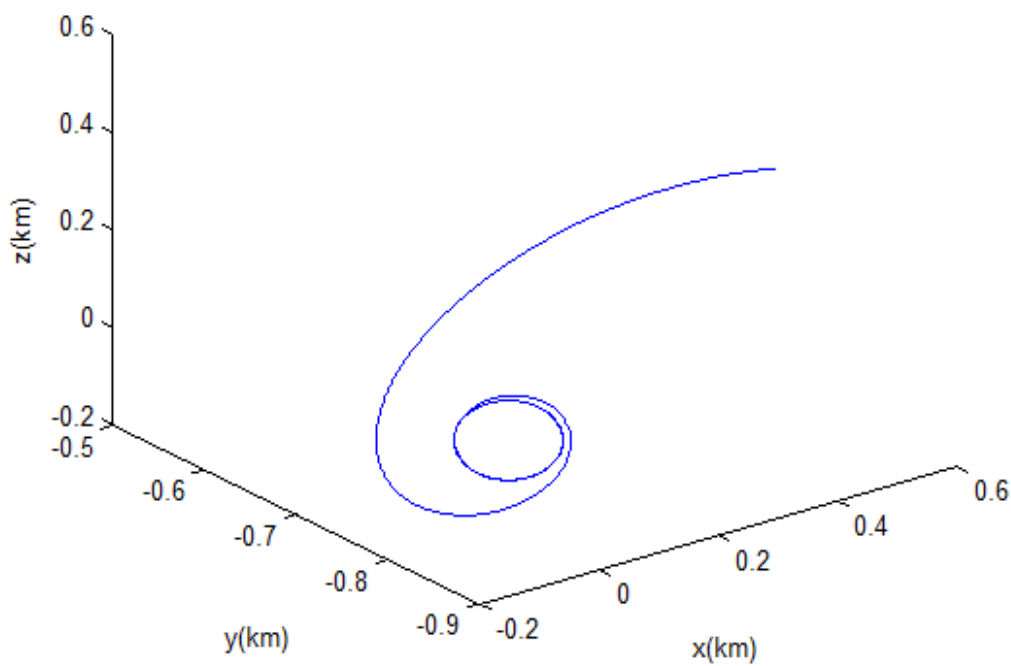

Figure 12. An Orbit near the Equilibrium Point E4

Figure 11 shows an almost periodic orbit near the equilibrium point E3. The flight time of the orbit is 4 days. There is only one family of periodic orbits near E3. Figure 12 shows an almost periodic orbit near the equilibrium point E4. The flight time of the orbit around the equilibrium point E4 is 4 days. There is only one family of periodic orbits near E4.

**6.4 Application to Asteroid 6489 Golevka**

The estimated bulk density of asteroid 6489 Golevka is 2.7 $\text{g}\cdot\text{cm}^{-3}$ (Mottola et al. 1997), and the rotation period of asteroid 6489 Golevka is 6.026h (Mottola et al. 1997). Table 7 shows the positions of the equilibrium points, which were calculated using Eq. (9). Table 8 shows the eigenvalues of the equilibrium points. E1 and E2 belong to Case 2, whereas E3 and E4 belong to Case 1.

Table 7 Positions of the Equilibrium Points around Asteroid 6489 Golevka

| Equilibrium Points | x (km) | y (km) | z (km) |
|---|---|---|---|
| E1 | 0.564128 | -0.0234156 | -0.002882 |
| E2 | -0.571527 | 0.035808 | -0.006081 |
| E3 | -0.021647 | 0.537470 | -0.001060 |
| E4 | -0.026365 | -0.546646 | -0.000182 |

Table 8 Eigenvalues of the Equilibrium Points around Asteroid 6489 Golevka

| $\times 10^{-3}$ | $\lambda_1$ | $\lambda_2$ | $\lambda_3$ | $\lambda_4$ | $\lambda_5$ | $\lambda_6$ |
|---|---|---|---|---|---|---|
| E1 | 0.125 | -0.125 | 0.297i | -0.297i | 0.309i | -0.309i |
| E2 | 0.181 | -0.181 | 0.304i | -0.304i | 0.329i | -0.329i |
| E3 | 0.207i | -0.207i | 0.216i | -0.216i | 0.280i | -0.280i |
| E4 | 0.185i | -0.185i | 0.213i | -0.213i | 0.297i | -0.297i |

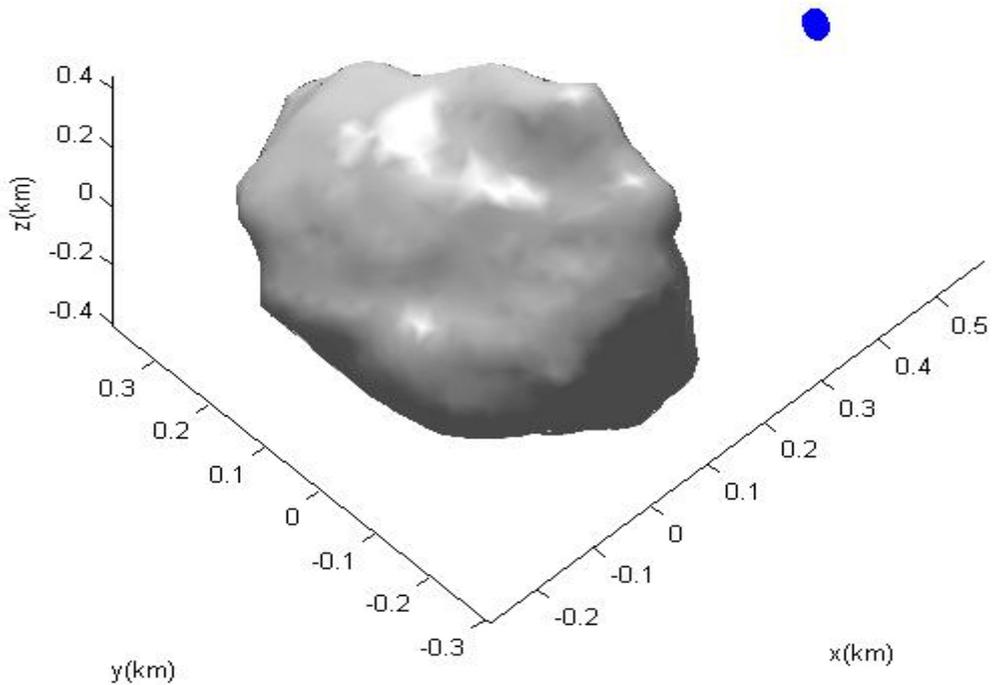

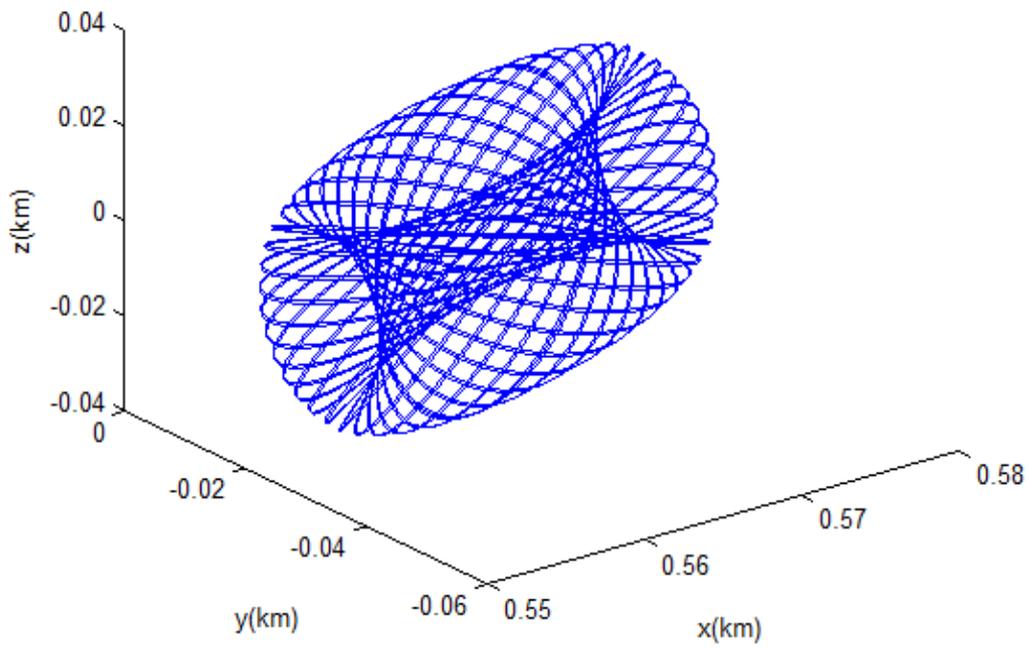

Figure 13.   A Quasi-Periodic Orbit near the Equilibrium Point E1

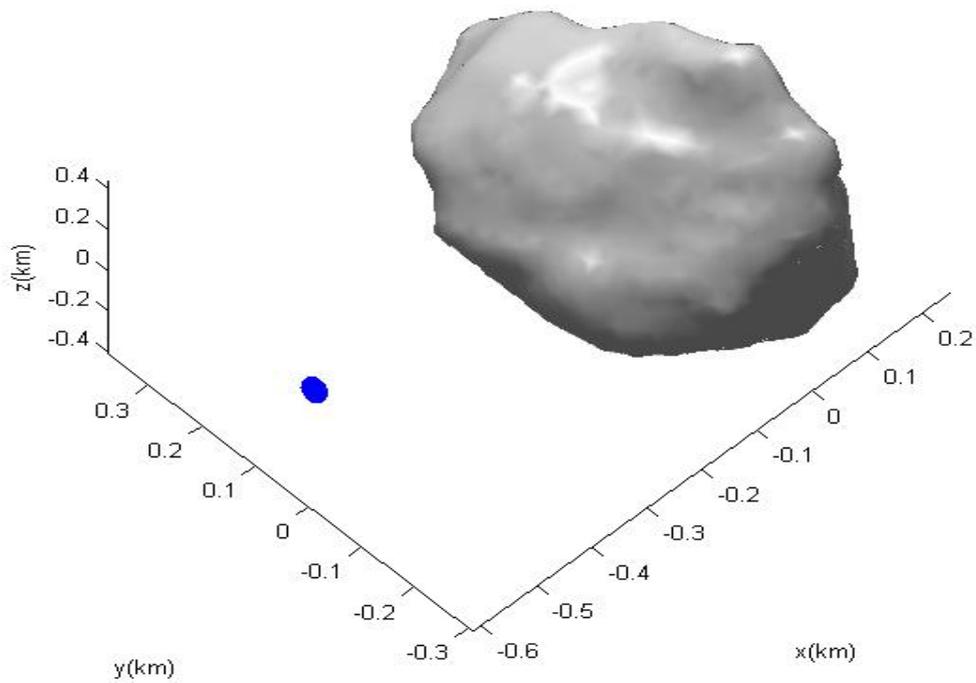

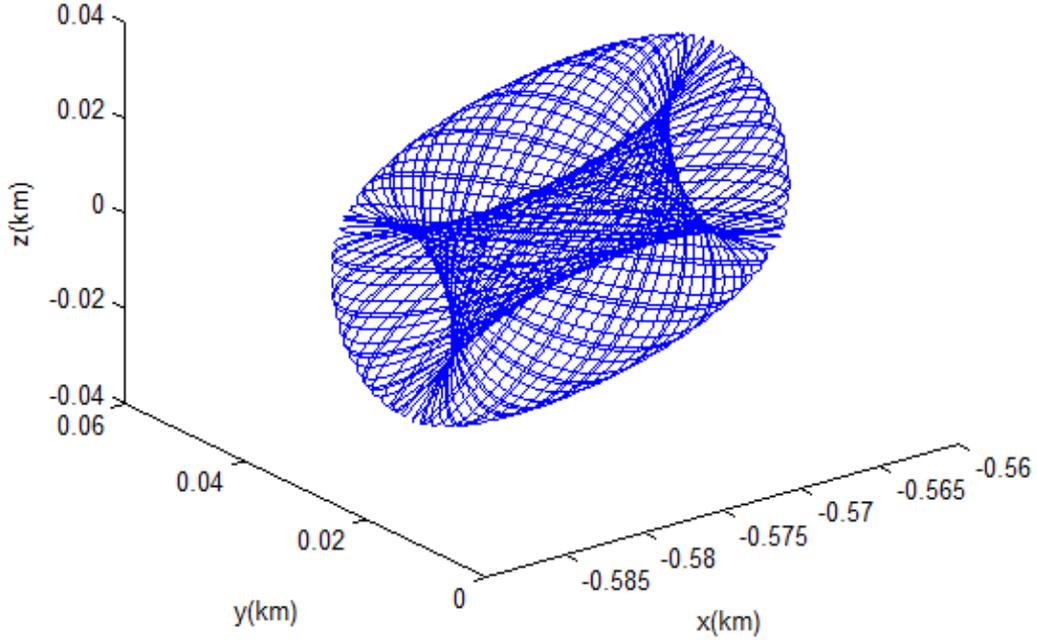

Figure 14.  A Quasi-Periodic Orbit near the Equilibrium Point E2

Figure 13 shows a quasi-periodic orbit near the equilibrium point E1, where the coefficients have the values $\begin{cases} C_{\xi1} = C_{\zeta1} = S_{\eta1} = C_{\eta2} = 0.01 \\ S_{\zeta2} = 0.02 \end{cases}$ and other coefficients being equal to zero. The flight time of the orbit is 4 days. E1 belongs to Case 2. There is one family of quasi-periodic orbits near E1, which is on the 2-dimensional tori $T^2$. The orbit in figure 13 belongs to the quasi-periodic orbital family. For the equilibrium point E2, figure 14 shows a quasi-periodic orbit near the equilibrium point E2, the coefficients have the values $\begin{cases} C_{\xi1} = C_{\zeta1} = S_{\eta1} = C_{\eta2} = 0.01 \\ S_{\zeta2} = 0.02 \end{cases}$ and other coefficients being equal to zero. The flight time of the orbit is 4 days. E2 also belongs to Case 2. There is one family of quasi-periodic orbits near E2, which is on the 2- dimensional tori $T^2$. The orbit belongs a quasi-periodic orbital family that is notably similar to that of E1.

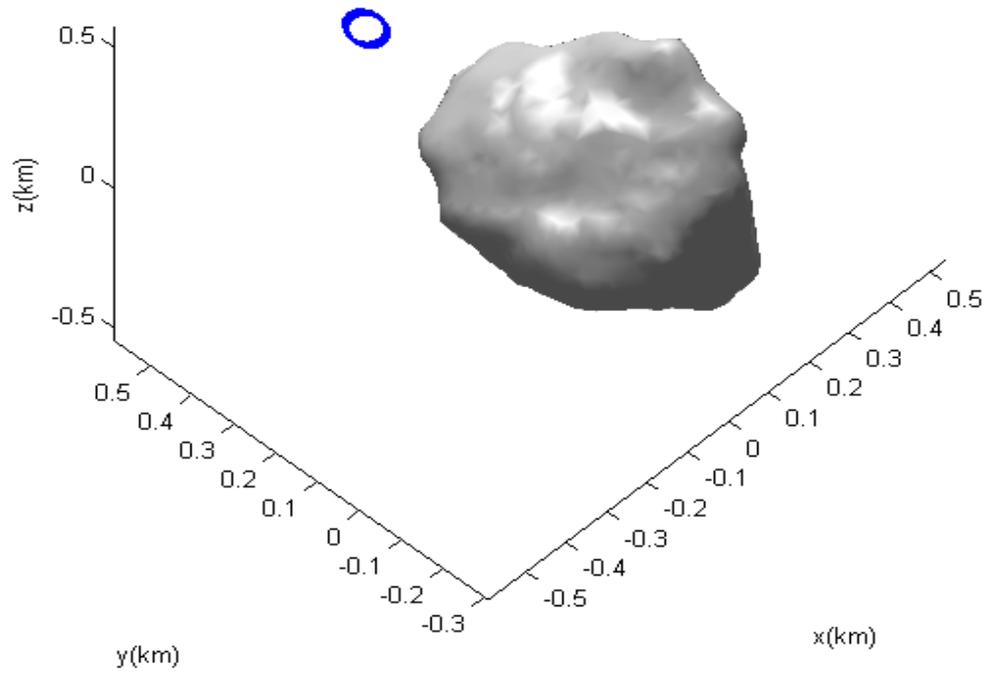

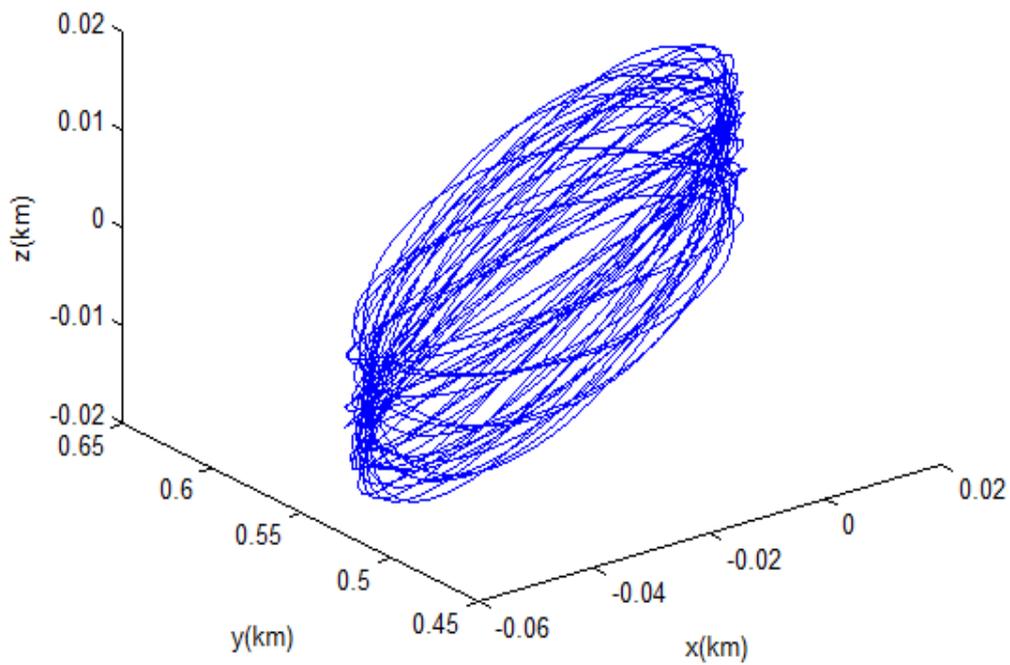

Figure 15. A Quasi-Periodic Orbit near the Equilibrium Point E3

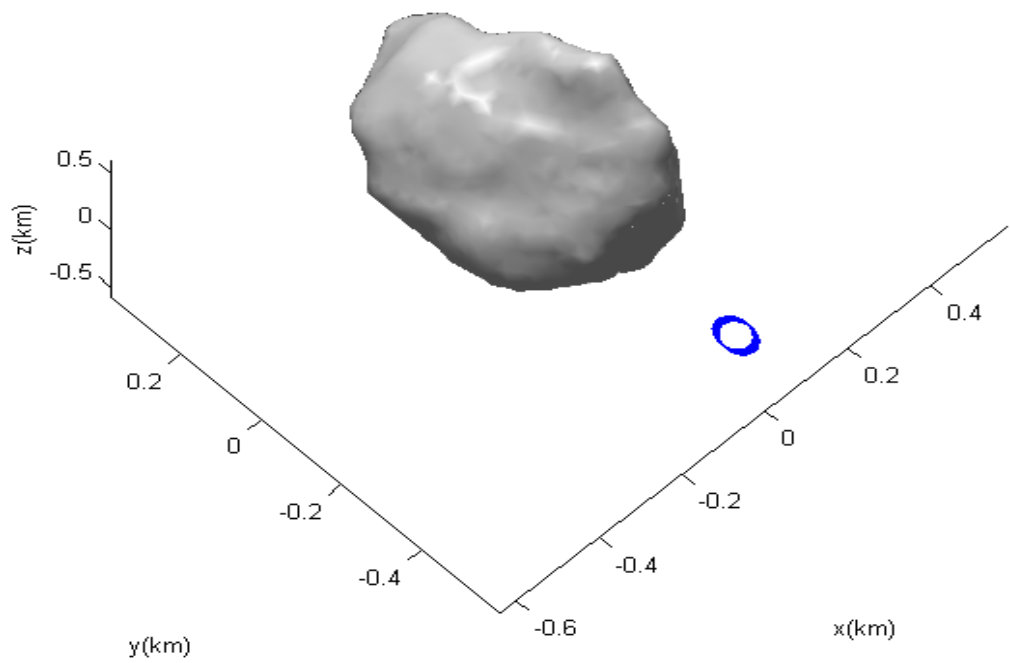

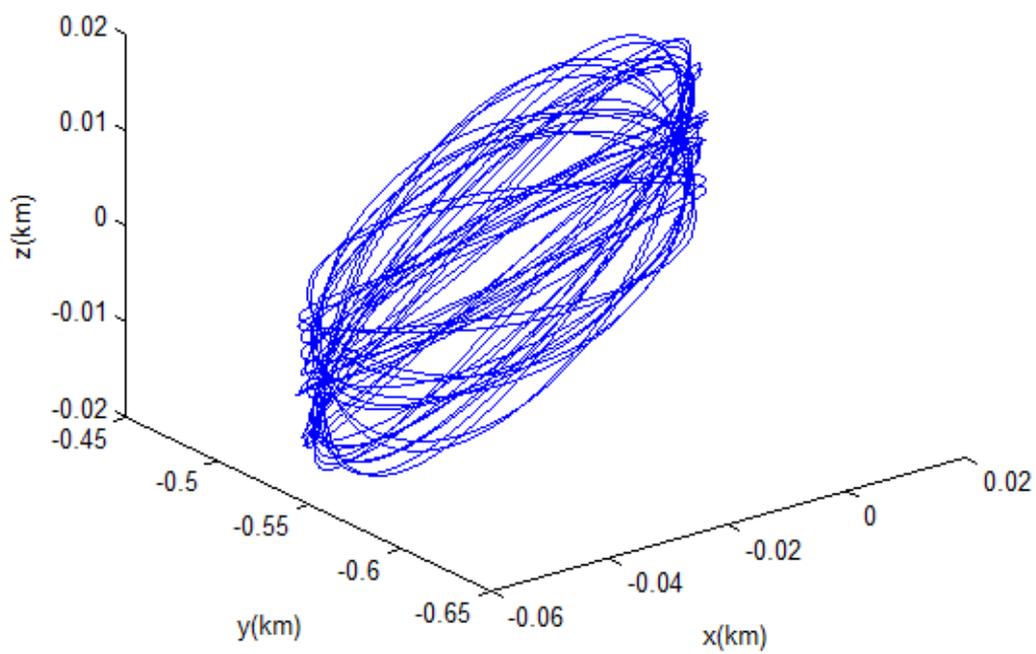

Figure 16.  A Quasi-Periodic Orbit near the Equilibrium Point E4

Figure 15 shows a quasi-periodic orbit near the equilibrium point E3, where the coefficients have the values $\begin{cases} C_{\xi 1} = 0.03 \\ S_{\eta 1} = 0.04 \\ C_{\zeta 1} = 0.01 \end{cases}$, $\begin{cases} C_{\eta 2} = 0.005 \\ S_{\zeta 2} = 0.006 \end{cases}$, $\begin{cases} C_{\eta 3} = 0.003 \\ S_{\zeta 3} = 0.002 \end{cases}$ and other coefficients being equal to zero. The flight time of the orbit is 12 days. E3 belongs to Case 1. There are four families of quasi-periodic orbits near E3, which are on the 3-dimensional tori $T^3$. The orbit in figure 15 belongs to a quasi-periodic orbital family. For the equilibrium point E4, figure 16 shows a quasi-periodic orbit near the equilibrium point E4, the coefficients have the values $\begin{cases} C_{\xi 1} = 0.03 \\ S_{\eta 1} = 0.04 \\ C_{\zeta 1} = 0.01 \end{cases}$, $\begin{cases} C_{\eta 2} = 0.005 \\ S_{\zeta 2} = 0.006 \end{cases}$, $\begin{cases} C_{\eta 3} = 0.003 \\ S_{\zeta 3} = 0.002 \end{cases}$ and other coefficients being equal to zero. The flight time of the orbit is 12 days. E4 also belongs to Case 1. There are four families of quasi-periodic orbits near E4, which are on the 3-dimensional tori $T^3$. The orbit belongs a quasi-periodic orbital family that is notably similar to that of E3.

**6.5 Discussion of Applications**

The theorems described in the previous sections are applied to asteroids 216 Kleopatra, 1620 Geographos, 4769 Castalia, and 6489 Golevka.

For the asteroid 216 Kleopatra, 1620 Geographos, and 4769 Castalia, the equilibrium points E1 and E2 belong to Case 2, whereas the equilibrium points E3 and E4 belong to Case 5. There are two families of periodic orbits and one family of quasi-periodic orbits on the central manifold near each of the equilibrium points E1 and E2. There is only one family of periodic orbits on the central manifold near each of the equilibrium points E3 and E4.

For the asteroid 6489 Golevka, E1 and E2 belong to Case 2, whereas E3 and E4 belong to Case 1. There are two families of periodic orbits and one family of quasi-periodic orbits on the central manifold near each of the equilibrium points E1 and E2. The equilibrium points E3 and E4 around asteroid 6489 Golevka are linearly stable. There are three families of periodic orbits and four families of quasi-periodic orbits on the central manifold near each of the equilibrium points E3 and E4.

## 7. Conclusions

The motion of a particle in the potential field of a rotating asteroid is studied, and the linearised equation of motion relative to the equilibrium points is presented. The characteristic equation of equilibrium points is presented and discussed. In addition, one sufficient condition and one necessary and sufficient condition for the stability of the equilibrium points in the potential field of a rotating asteroid are provided.

Considering the orbit of the particle near the asteroid, we link the orbit and the geodesic of a smooth manifold with a new metric. The metric does not have a positive definite quadratic form. Then, we classify the equilibrium points into eight cases using the eigenvalues. The structure of the submanifold near the equilibrium point is related to the eigenvalues. A theorem is presented to describe the structure of the submanifold as well as the stable and unstable behaviours of the particle near the equilibrium points.

Near the linearly stable equilibrium point, there are four families of quasi-periodic orbits, which are on the k-dimensional tori $T^k (k=2,3)$. There are neither asymptotically stable manifold nor asymptotically unstable manifold near the linearly stable equilibrium point. The structures of the submanifold and the subspace of the linearly stable equilibrium point are different from those of the linear unstable equilibrium point.

Near the non-resonant unstable equilibrium points, the dimension of the unstable manifold is greater than zero. The periodic orbit, the quasi-periodic orbit, etc., have been studied near the non-resonant unstable equilibrium points. Near the resonant equilibrium points, there is at least one family of periodic orbits. The dimension of the resonant manifold is greater than four.

The theory is applied to asteroids 216 Kleopatra, 1620 Geographos, 4769 Castalia, and 6489 Golevka. For the asteroid 216 Kleopatra, 1620 Geographos, and 4769 Castalia, the equilibrium points are unstable, which are denoted as E1, E2, E3, and E4, two of them belong to Case 2 while the other two of them belong to Case 5. For the asteroid 6489 Golevka, two equilibrium points belong to Case 2; whereas the other two equilibrium points belong to Case 1, i.e. they are linearly stable.

**Acknowledgements**

This research was supported by the National Basic Research Program of China (973 Program, 2012CB720000) and the National Natural Science Foundation of China (No. 11072122).